\documentclass[aps,twocolumn,showpacs,preprintnumbers,amsmath,amssymb,nofootinbib,superscriptaddress,showkeys,floatfix,superscriptaddress]{revtex4}

\usepackage{epsfig}
\usepackage{graphicx}

\def\u{{\bf u}} \def\U{{\bf U}} \def\S{{\bf S}} \def\h{{\bf h}}
 \def\E{{\bf 1}}

\begin{document}

\title{Renormalization of NN Interaction with Lorentz-invariant Chiral 
Two Pion Exchange.} 

\author{R. Higa}
\email{higa@itkp.uni-bonn.de}
\affiliation{ Universit{\"a}t Bonn, HISKP (Theorie),
                        Nussallee 14-16, 53115 Bonn, Germany}
\affiliation{ Jefferson Laboratory,
              12000 Jefferson Avenue, Newport News, VA 23606 USA}
  \author{M. Pav\'on Valderrama}\email{mpavon@ugr.es}
  \affiliation{Departamento de F\'{\i}sica At\'omica, Molecular y
  Nuclear, Universidad de Granada, E-18071 Granada, Spain.}
\affiliation{The
       H. Niewodnicza\'nski Institute of Nuclear Physics,  PL-31342
       Krak\'ow, Poland}
  \author{E. Ruiz Arriola} \email{earriola@ugr.es}
  \affiliation{Departamento de F\'{\i}sica At\'omica, Molecular y
  Nuclear, Universidad de
  Granada, E-18071 Granada, Spain.}

\date{\today}

\begin{abstract} 
\rule{0ex}{3ex} The renormalization of the NN interaction with the
Chiral Two Pion Exchange Potential computed using Lorentz-invariant baryon
chiral perturbation theory is considered. The short distance
singularity reduces the number of counter-terms to about a half as
those in the heavy baryon expansion. Phase shifts and deuteron
properties are evaluated and a general overall agreement is
observed. \\
\end{abstract}
\pacs{03.65.Nk,11.10.Gh,13.75.Cs,21.30.Fe,21.45.+v} 
\keywords{NN interaction, Two Pion Exchange, Renormalization, Relativity}
\maketitle



\section{Introduction}

At long distances the Nucleon-Nucleon (NN) interaction can be written
as (see e.g. \cite{Brown:76} for a review)
\begin{eqnarray} 
{\bf V}(r) = {\bf V}_{1\pi} (r)+ {\bf V}_{2\pi} (r) + \dots 
\label{eq:full_pot} 
\end{eqnarray} 
where, $ {\bf V}_{1\pi} (r)$ and $ {\bf V}_{2\pi} (r) $ represent the
One Pion Exchange (OPE) and the Two Pion Exchange (TPE) contributions
to the potential, respectively.  Up to power corrections in the
distance $r$, one has $ {\bf V}_{n \pi} (r) = {\cal O} ( e^{-n m_\pi
r}) $. Such an expansion makes sense, since there is a clear scale
separation at long distances.  Actually, the omitted terms in
Eq.~(\ref{eq:full_pot}) represent contributions whose ranges are shorter
than $1/(2 m_\pi) \sim 0.7 {\rm fm}$. These include three and higher
pion exchanges, correlated meson exchanges,
etc~\cite{Machleidt:1987hj}. In momentum space, the expansion of
Eq.~(\ref{eq:full_pot}) parallels an expansion on leading low momentum
singularities, rather than a naive low momentum expansion. The
systematic and model independent determination of those potentials was
suggested several years
ago~\cite{Weinberg:1990rz,Ordonez:1992xp,Weinberg:1991um,Ordonez:1995rz}
and pursued by many
others~\cite{Rijken:1995pu,Kaiser:1997mw,Friar:1999sj,Rentmeester:1999vw,Entem:2002sf}
(for some reviews emphasizing different viewpoints see
e.g.~\cite{Bedaque:2002mn,Phillips:2002da,Epelbaum:2005pn,Rho:2006tx,
RuizArriola:2006hc} and references therein). However, in any scheme
the potentials in Eq.~(\ref{eq:full_pot}) become singular at short
distances, so one must truncate or renormalize the potential in a
physically meaningful way in order to predict finite and unique phase
shifts and deuteron properties. This has been a subject of much debate
and controversy in recent times and we refer the interested reader to
the literature for further details~\cite{Epelbaum:1999dj,Entem:2001cg,
Rentmeester:1999vw,Beane:2001bc,Entem:2003ft,Epelbaum:2004fk,PavonValderrama:2004nb,Birse:2003nz,Nogga:2005hy,
PavonValderrama:2005gu,Valderrama:2005wv,Birse:2005um,
PavonValderrama:2005uj,Erratum,Epelbaum:2006pt,Birse:2007sx}.

In previous works by two of us (MPV and 
ERA)~\cite{PavonValderrama:2005gu,Valderrama:2005wv,PavonValderrama:2005uj,Erratum} 
the renormalization of NN potentials was studied using chiral potentials 
based on the heavy baryon formalism 
(HB-$\chi$PT)~\cite{Kaiser:1997mw,Rentmeester:1999vw}. 
In the present paper, we extend those ideas to the case where
Lorentz-invariant chiral potentials are used instead. In this case,
the relativistic framework of baryon chiral perturbation theory
(RB-$\chi$PT) proposed by Becher and
Leutwyler~\cite{Becher:1999he,Becher:2001hv} is employed in the
construction of the two-pion exchange (TPE) component of the NN
interaction.  The remarkable difference between HB- and RB-TPE
potentials lies on the long distance
behavior~\cite{Higa:2003jk,Higa:2003sz,Higa:2004cr,Higa:2005ip}, due to
the analytic structure of the $\pi N$ scattering amplitude in the low
energy region where the Mandelstam variable $t$ is close to
$4m_{\pi}^2$, $m_{\pi}$ being the pion mass. Actually, one appealing
feature of the RB-TPE potentials is that the long-distance two pion
effects are correctly described, so that important contributions at
the exponential level $\sim e^{- 2 m_\pi r}$ are properly re-summed,
unlike its heavy baryon counterpart.  In this work we are also
interested on its different short distance behavior, which plays an
important role on the renormalization program of the NN interaction
developed in
Refs.~\cite{PavonValderrama:2005gu,Valderrama:2005wv,PavonValderrama:2005uj,Erratum}. We
disregard, however, explicit $\Delta$'s (see e.g. \cite{Kaiser:1998wa})
and other intermediate state contributions, assuming that those degrees of 
freedom have been integrated out.

As we have already mentioned, the calculation of scattering and bound
state properties requires specifying the NN potential at short distances, 
which turns out to be highly singular for the Lorentz-invariant
case. Actually, a crucial issue in the present context regards the
number of necessary counter-terms required by the renormalizability of
the S-matrix. In the single channel situation, the results found in
Refs.~\cite{PavonValderrama:2005gu,Valderrama:2005wv,PavonValderrama:2005uj,Erratum}
in configuration space can be summarized as follows. If the potential
is a regular one, i.e., $r^2 |U(r)| < \infty $, there is freedom to
{\it choose} between the regular and irregular solution of the
corresponding Schr\"odinger equation. In the first case the scattering
length is predicted, while in the second case the scattering length
becomes an input of the calculation. Singular potentials fulfill $r^2
|U(r)| \to \infty $ and do not allow this choice. For repulsive
singular potentials [$r^2 U(r) \to \infty $] the scattering length
is {\it predicted} while for attractive singular potentials [$r^2 U(r)
\to -\infty $] the scattering length {\it must} be given. The case
$r^2 U(r) \to g$ is very special and, for $g< -1/4$, yields ultraviolet
limit-cycles~\cite{Beane:2000wh,PavonValderrama:2004nb,Valderrama:2007nu}. For
coupled channels one must diagonalize first the coupled channel
potentials and apply the single channel rules to the outgoing
eigenpotentials.

In order to avoid any possible misunderstanding we hasten to emphasize
that our use of the word relativistic is in a narrow sense; we are
only disregarding a naive heavy baryon expansion of the virtual
nucleon states in the calculation of the potential and hence taking
into account important anomalous thresholds
singularities~\cite{Becher:1999he,Becher:2001hv}.  This is {\it not
the same} as providing a fully relativistic quantum field theoretical
solution to the the two-body problem by say, solving a Bethe-Salpeter
equation or any two body relativistic equation. This has always been a
problem rooted in the non-perturbative divorce between crossing and
unitarity in few body calculations for which the present paper has
nothing to say.

The paper is organized as follows.  In Sec.~\ref{sec:form} we give an
overview of our formalism already used in
Ref.~\cite{PavonValderrama:2005gu,Valderrama:2005wv,PavonValderrama:2005uj,Erratum}.
The key aspects on the derivation of the Lorentz-invariant TPE
potential are briefly mentioned and the main differences with respect
to the heavy baryon formalism are highlighted in Sec.~\ref{sec:relat}.
The deuteron bound state is discussed in Sec.~\ref{sec:deuteron}. Our
predictions for phase shifts are displayed in
Sec.~\ref{sec:phases}. Finally, in Sec.~\ref{sec:concl} we draw our
conclusions.

\begin{table}
\caption{\label{tab:table1} Sets of chiral coefficients
considered in this work.}
\begin{tabular}{|c|c|c|c|c|}
\hline  Set & Source & $c_1 ({\rm GeV}^{-1}) $ & $c_3 ({\rm GeV}^{-1}) $ & $c_4
({\rm GeV}^{-1}) $ \\ \hline
Set I & $\pi N$~\cite{Buettiker:1999ap}  & -0.81  & -4.69    & 3.40   \\ 
Set II & $NN $~\cite{Rentmeester:1999vw} 
& -0.76  & -5.08   & 4.70  \\ 
Set III &  $NN $~\cite{Epelbaum:2003xx}  
& -0.81  & -3.40    & 3.40   \\ 
Set IV &  $NN$~\cite{Entem:2003ft} & -0.81  & -3.20   & 5.40  \\ 
Set $\eta$ &  This work & -0.81  & -3.80   & 4.50  \\ 
\hline
\end{tabular}
\end{table}

\begin{table*}
\caption{\label{tab:table2} The number of independent parameters
(counter-terms) for the Lorentz-invariant baryon expansion potential (RB)
different orders of approximation of the heavy baryon expansion (HB)
potential. The scattering lengths are in ${\rm fm}^{l+l'+1}$ and are
taken from NijmII and Reid93 potentials~\cite{Stoks:1994wp} in
Ref.~\cite{Valderrama:2005ku}. We use the (SYM-nuclear bar)
convention, Eq.~(\ref{eq:phase-thres}).  The stars (*) mean that the
behaviour is very dependent on the chosen set of chiral couplings.}
\begin{tabular}{|c|c|c|c|c|c|c|}
\hline 
Wave  & $\alpha$ NijmII (Reid93) & OPE & HB-NLO & HB-NNLO  & HB-NNLO & RB-TPE \\ 
      &   &   &   & Set I, II \& III & SetIV &  \\ 
\hline 
\hline $^1S_0 $ & -23.727(-23.735) & Input & Input & Input & Input & Input  \\
\hline $^3P_0 $ & -2.468(-2.469) &  Input & --- & Input & ---  & ---(*) \\
\hline
\hline 
$^1P_1 $ & 2.797(2.736)  & --- & --- & Input & --- & --- \\
\hline $^3P_1 $ & 1.529(1.530) & --- & Input & Input & Input & Input \\
\hline 
$^3S_1 $ & 5.418(5.422) & Input  &  ---  & Input & Input & Input \\
$^3D_1 $ & 6.505(6.453) &  --- &  --- & Input & Input  & --- \\
$E_1 $ & 1.647(1.645)   & ---   & ---  & Input & Input  & --- \\
\hline 
\hline 
$^1D_2 $ & -1.389(-1.377) & ---  & Input & Input & Input  & Input \\
\hline $^3D_2 $ & -7.405(-7.411) & Input  &  Input & Input & Input & Input \\
\hline $^3P_2 $ & -0.2844(-0.2892) & Input  &  Input & Input & Input & Input \\
$^3F_2 $ & -0.9763(-0.9698) & --- & --- & Input & ---  & ---(*) \\
$E_2 $ & 1.609(1.600) & --- & ---  & Input & --- & ---(*) \\
\hline
\hline 
$^1F_3 $ & 8.383(8.365) & ---  & --- & Input & ---  & Input \\
\hline $^3F_3 $ & 2.703(2.686) & --- &  Input & Input & Input & Input \\
\hline 
$^3D_3 $ & -0.1449(-0.1770) & Input  & --- & Input & Input & Input  \\
$^3G_3 $ & 4.880(4.874) & ---  & --- & Input & Input  & ---  \\
$E_3 $ & -9.695(-9.683) & --- & --- & Input & Input  & --- \\
\hline 
\hline 
$^1G_4 $ & -3.229(-3.210) & ---  & Input  & Input & Input & Input \\
\hline $^3G_4 $ & -19.17(-19.14) & Input  & Input  & Input & Input & Input  \\
\hline 
$^3F_4 $ & -0.01045(-0.01053) & Input  & Input  & Input & Input & Input \\
$^3H_4 $ & -1.250(-1.240) & --- & --- & Input & --- & --- (*) \\
$E_4 $ & 3.609(3.586) & --- & --- & Input & ---  & --- (*) \\
\hline
\hline 
$^1H_5 $ & 28.61(28.57) & --- & --- & Input & --- & Input \\
\hline 
$^3H_5 $ & 6.128(6.082) & --- & Input  & Input & Input & Input  \\
\hline 
$^3G_5 $ & -0.0090(-0.010) & Input  & --- & Input & Input  & Input  \\
$^3I_5 $ & 10.68(10.66) & --- &--- & Input & Input  & ---  \\
$E_5 $ & -31.34(-31.29) & --- & --- & Input & Input  & --- \\
\hline
\end{tabular}
\end{table*}

\section{Formalism}
\label{sec:form}

Along the lines of
Ref.~\cite{PavonValderrama:2005gu,Valderrama:2005wv,PavonValderrama:2005uj,Erratum} we solve the
coupled channel Schr\"odinger equation in configuration space for the
relative motion, which in compact notation reads
\begin{eqnarray}
-\u '' (r) + \left[ \U (r) + \frac{{\bf l}^2}{r^2} \right] \u (r) =
 k^2 \u (r) \, . 
\label{eq:sch_cp} 
\end{eqnarray} 
The coupled channel matrix reduced potential is defined as usual, 
$\U (r)= 2 \mu_{np} {\bf V}(r)$, where $\mu_{np}=M_p M_n /(M_p+M_n)$ 
is the reduced proton-neutron mass. For $j> 0$, $\U (r)$ can be written as
\begin{eqnarray}
\U^{0j} (r) &=& U_{jj}^{0j} \, , \nonumber \\ \\   
\U^{1j} (r) &=& \begin{pmatrix}  U_{j-1,j-1}^{1j} (r) & 0 &
U_{j-1,j+1}^{1j} (r) \\ 0 & U_{jj}^{1j} (r) & 0 \\ U_{j-1,j+1}^{1j}
(r) & 0 & U_{j+1,j+1}^{1j} (r)   \end{pmatrix} \, . \nonumber  
\end{eqnarray} 
In Eq.~(\ref{eq:sch_cp}) $ {\bf l}^2 = {\rm diag} ( l_1 (l_1+1),
\dots, l_N (l_N +1) )$ is the angular momentum, $\u(r)$ is the reduced
matrix wave function and $k$ the C.M. momentum. In our case, $N=1$ for
the spin singlet channel ($l=j$) and $ N=3 $ for the spin triplet
channel, with $l_1=j-1$, $l_2=j$ and $l_3=j+1$. The potentials used in
this paper were obtained in Refs.~\cite{Higa:2003jk,Higa:2003sz,
Higa:2004cr}, in coordinate space. We outline the main issues of
this potential in Sec.~\ref{sec:relat}.

\subsection{Long distance behaviour} 

At long distances, we assume the usual asymptotic normalization
condition
\begin{eqnarray}
\u (r)  \to \hat \h^{(-)} (r) - \hat \h^{(+)} (r) \S \, ,
\label{eq:asym}
\end{eqnarray} 
with $\S$ the  coupled channel unitary S-matrix. The
corresponding outgoing and incoming free spherical waves are given by
\begin{eqnarray}
\hat \h^{(\pm)} (r) &=& {\rm diag} ( \hat h^\pm_{l_1} ( k r) , \dots ,
\hat h^\pm_{l_N} (k r) ) \, ,
\end{eqnarray} 
with $ \hat h^{\pm}_l ( x) $ the reduced Hankel functions of order
$l$, $ \hat h_l^{\pm} (x) = x H_{l+1/2}^{\pm} (x) $ ( $ \hat h_0^{\pm}
(x) = e^{ \pm i x}$ ), and satisfy the free Schr\"odinger's equation
for a free particle. 

The spin singlet state ($s=0$) is an un-coupled state
\begin{eqnarray}
S_{jj}^{0j} = e^{ 2 i \delta_{j}^{0j} } \, ,
\end{eqnarray}
while the spin triplet state ($s=1$) comprises one un-coupled $ l=j$
state
\begin{eqnarray}
S_{jj}^{1j} &=& e^{  2 i \delta_{j}^{1j} } \, ,
\end{eqnarray}
and the two channel coupled $l,l'=j \pm 1$ states for which we use
Stapp-Ypsilantis-Metropolis (SYM or Nuclear bar)~\cite{stapp}
parameterization 
\begin{eqnarray}
S^{1j} &=& \left( \begin{array}{cc} S_{j-1 \,
j-1}^{1j} & S_{j-1 \, j+1}^{1j} \\ S_{j+1 \, j-1}^{1j} & S_{j+1 \,
j+1}^{1j}
\end{array} \right) \nonumber  \\ &=& \left( \begin{array}{cc} \cos{(2
\bar \epsilon_j)} e^{2 i \bar \delta^{1j}_{j-1}} & i \sin{(2 \bar
\epsilon_j)} e^{i (\bar \delta^{1j}_{j-1} +\bar \delta^{1j}_{j+1})} \\
i \sin{(2 \bar \epsilon_j)} e^{i (\bar \delta^{1j}_{j-1} + \bar
\delta^{1j}_{j+1})} & \cos{(2 \bar \epsilon_j)} e^{2 i \bar
\delta^{1j}_{j+1}} \end{array} \right) \nonumber
\end{eqnarray}
In the present paper zero energy scattering parameters play an
essential role, since they are often used (see below) as input
parameters in the calculation of phase shifts. Due to unitarity of the
S-matrix in the low energy limit, $ k\to 0$ we have
\begin{eqnarray}
\left(\S - \E \right)_{l',l}=- 2 {\rm i} \alpha_{l', l}  k^{l'+l+1} +
\dots   \, ,  
\end{eqnarray} 
with $\alpha_{l' l} $ the (hermitian) scattering length
matrix. The threshold behaviour of the SYM phases is 
\begin{eqnarray} 
\bar \delta^{1j}_{j-1} &\to&  - \bar \alpha^{1j}_{j-1} k^{2j-1} \, , \\ 
\bar \delta^{1j}_{j+1} &\to&  - \bar \alpha^{1j}_{j+1} k^{2j+3} \, , \\ 
\bar \epsilon_j &\to&  - \bar \alpha^{1j}_{j} k^{2j+1} \, . 
\label{eq:phase-thres}
\end{eqnarray} 

\subsection{Short distance behaviour} 

The form of the wave functions at the origin is uniquely determined by
the form of the potential at short distances (see
e.g.~\cite{Case:1950an,Frank:1971xx} for the case of one channel and
\cite{PavonValderrama:2005gu,Valderrama:2005wv,PavonValderrama:2005uj,Erratum} for coupled
channels). For the Lorentz-invariant chiral NN potential , one
has
\begin{eqnarray}
\U_{2 \pi} (r) &\to& \frac{M {\bf C}_{7}}{r^7} \, , \nonumber 
\label{eq:} 
\end{eqnarray}
which is a relativistic Van der Waals type force (see
e.g.~\cite{Feinberg:1989ps} for the electromagnetic case). Note that
this short distance behaviour without an $1/M$ expansion 
is at variance with the non-relativistic $1/r^5$ and $1/r^6$ in the
standard Weinberg counting based on the HB chiral expansion. In the 
latter, the expansion around the limit $M\to \infty$ is built in the 
formalism, leading to a different behavior at $r \to 0$. 

For a potential diverging at the origin as an inverse power law one has
\begin{eqnarray}
\U (r) \to \frac{M {\bf C}_n}{r^n} \, , 
\label{eq:singular} 
\end{eqnarray} 
with ${\bf C}_n$ a matrix of generalized Van der Waals coefficients
and $n > 2$. One diagonalizes the matrix ${\bf C}_n $ by a constant
unitary transformation, ${\bf G}$, yielding
\begin{eqnarray}
M {\bf C}_n = {\bf G} \, {\rm diag} ( \pm R_1^{n-2}, \dots , \pm R_N^{n-2} ) \,
{\bf G}^{-1} \, , 
\end{eqnarray} 
with $ R_i$ constants with length dimension. The plus sign corresponds
to the case with a positive eigenvalue (attractive) and the minus
sign to the case of a negative eigenvalue (repulsive). Then, at short
distances one has the solutions
\begin{eqnarray}
\u (r) \to {\bf G} \begin{pmatrix}  u_{1,\pm} (r) \cr \cdots \\
u_{N,\pm} (r)
\end{pmatrix} \, ,
\label{eq:eigen_wf}
\end{eqnarray} 
where for the attractive and repulsive cases one has 
\begin{eqnarray}
u_{i,-} (r) &\to & C_{i,-} \left(\frac{r}{R_i}\right)^{n/4} \sin\left[
\frac{2}{n-2} \left(\frac{R_i}{r}\right)^{\frac{n}2-1} + \varphi_i
\right] \, ,\nonumber \\ \label{eq:uA} \\ u_{i,+} (r) & \to & C_{i,+}
\left(\frac{r}{R_i}\right)^{n/4} \exp \left[- \frac{2}{n-2}
\left(\frac{R_i}{r}\right)^{\frac{n}2-1} \right] \, , \nonumber \\ \label{eq:uR}
\label{eq:short_wf}
\end{eqnarray} 
respectively. Here, $\varphi_i$ are arbitrary short distance phases which
in general depend on the energy. There are as many short distance
phases as short distance attractive eigenpotentials. Orthogonality of
the wave functions at the origin yield the relation
\begin{eqnarray}
\sum_{i=1}^N \left[ {u_{k,i}}^* u_{p,i}'- {u_{k,i}'}^* 
u_{p,i} \right]\Big|_{r=0} = 
\sum_{i=1}^A \cos(\varphi_i (k) - \varphi_i(p) ) \, , \nonumber \\ 
\end{eqnarray} 
where $A \le N$ is the number of the short distance attractive
eigenpotentials.  Details on the numerical implementation of these
short distance boundary conditions can be looked up in
Refs.~\cite{PavonValderrama:2005gu,Valderrama:2005wv,PavonValderrama:2005uj,Erratum}. 

\subsection{Numerical parameters}

The Lorentz-invariant chiral TPE potential is specified by the pion
weak decay constant $f_\pi$, the nucleon axial coupling constant
$g_A$, the nucleon mass $M_N$ and the pion mass $m_\pi$. In addition,
at the level of approximation that we are working, it is enough to
consider the low energy constants $c_1$, $c_3$ and $c_4$ which
characterize $\pi N $ scattering. The corresponding RB potential is
specified by the same parameters at N$^3$LO in the HB chiral
expansion.

In our numerical calculations one takes $f_\pi=92.4 {\rm MeV}$,
$m_\pi=138.03 {\rm MeV}$, $ 2 \mu_{np}= M_N = 2M_p M_n /(M_p+M_n) =
938.918 {\rm MeV}$, $ g_A =1.29 $ in the OPE piece to account for the
Goldberger-Treimann discrepancy and $g_A=1.26 $ in the TPE piece of
the potential. The corresponding pion nucleon coupling constant takes
then the value $ g_{\pi NN}=13.083$, according to the Nijmegen phase
shift analysis of NN scattering~\cite{deSwart:1997ep}. The values of
the coefficients $c_1$, $c_3$ and $c_4$ used along this paper can be
looked up in Table~\ref{tab:table1}. There we list several Sets which
have been proposed in the
literature~\cite{Buettiker:1999ap,Rentmeester:1999vw,
Epelbaum:2003xx,Entem:2003ft} as well as the one which will be used in
the present work based on our analysis of deuteron properties below.

Renormalization requires fixing some low energy parameters while
removing the cut-off. We take the values from the high quality
potentials~\cite{Stoks:1993tb,Stoks:1994wp} as have been obtained in
Ref.~\cite{Valderrama:2005ku} for the NijmII and Reid93 versions. We
will use the Nijm II values for definiteness. As mentioned earlier,
the number of independent parameters or counter-terms requires a study
of the attractive/repulsive nature of the potential at short
distances. The result of such an analysis for all channels considered
in this work is summarized in table~\ref{tab:table2} for the different
parameter sets. We also list the scattering lengths in all partial
waves as determined in our previous work~\cite{Valderrama:2005ku}.

\section{Lorentz-invariant two-pion exchange}
\label{sec:relat}

\begin{figure}[!ht]
  \epsfig{figure=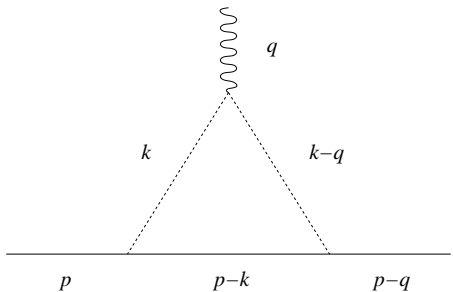, height=1.5in}
  \caption{The triangle diagram, which cannot be reproduced by the usual 
heavy baryon expansion 
close to $t=4m_{\pi}^2$. The solid, dashed, and wiggly lines represent,
respectively, the nucleon, the pions, and an external scalar source.}
\label{fig:triangle}
\end{figure}

A series of
papers~\cite{Higa:2003jk,Higa:2003sz,Higa:2004cr,Higa:2005ip} was
devoted to the construction of the TPE component of the NN
interaction, based on the Lorentz-invariant formulation of baryon
chiral perturbation theory proposed by Becher and
Leutwyler~\cite{Becher:1999he,Becher:2001hv} in their study of the
$\pi N$ system. These authors showed that it is possible to obtain a
consistent power counting in a theory with a heavy particle without
resorting to an integration of the heavy degrees of freedom and
expansion around the limit of infinitely heavy baryon
(HB-$\chi$PT). The latter results can be recovered from the
Lorentz-invariant formulation through an expansion in
$1/m_N$. However, this procedure destroys the correct analytic
behavior of the amplitude near the low energy region close to
$t=4m_{\pi}^2$.  The underlying reason comes from the anomalous
threshold of the triangle graph~\cite{Becher:1999he}
(Fig.\ref{fig:triangle}) right below threshold,
$t=4m_{\pi}^2-m_{\pi}^4/m_N^2$. In the heavy baryon limit this
singularity is ignored (as it collapses to the normal threshold) and
any $1/m_N$ expansion of the triangle loop around this region will
fail to converge. Note that the same triangle integral also appears in
the TPE potential, with two pseudo-vector vertices in one nucleon and
a Weinberg-Tomozawa seagull term on the other.

In order to illustrate the problem let us consider the spectral 
representation of the triangle graph, 

\begin{equation}
\gamma(t)=\frac{1}{\pi}\int_{4m_{\pi}^2}^{\infty}\frac{dt'}{(t'-t)}
\,\mbox{Im}\gamma(t')\;,
\end{equation}

\noindent
where

\begin{eqnarray}
\mbox{Im}\gamma(t') &\!=\!& 
\frac{\theta(t'\!-\!4m_{\pi}^2)}{16\pi m_N\sqrt{t'}}\,
\arctan\frac{2m_N\sqrt{t'\!-\!4m_{\pi}^2}}{t'\!-\!2m_{\pi}^2}\,. 
\end{eqnarray}

\noindent In HB-$\chi$PT the argument 
$x=2m_N\sqrt{t'-4m_{\pi}^2}/(t'-2m_{\pi}^2)$ is assumed to be of order 
$q^{-1}$, yielding the expansion 
$\arctan x= \pi/2 - 1/x + 1/3x^3+\cdots $. The first two terms read 

\begin{eqnarray}
&&\gamma(-q^2)|_{HB}=
\nonumber\\&&
\frac{1}{16\pi^2m_N m_{\pi}}\int_{4m_{\pi}^2}^{\infty}
\frac{dt'}{(t'+q^2)}\,\frac{1}{\sqrt{t'}}
\left[\frac{\pi}{2} - \frac{(t'-2m_{\pi}^2)}{2 m_N 
\sqrt{t'\!-\!4m_{\pi}^2}}\right]
\nonumber\\
&&\!=\!\frac{1}{16\pi^2m_N m_{\pi}}\left[2\pi m_{\pi}A(q)
\!+\!\frac{m_{\pi}}{m_N}
\frac{(2m_{\pi}^2\!+\!q^2)}{(4m_{\pi}^2\!+\!q^2)}L(q)\right],
\end{eqnarray}

\noindent where $q=|{\mbox{\boldmath $q$}}|$, and $L(q)$ and $A(q)$ are
the usual HB loop functions,

\begin{eqnarray}
&&L(q)=\frac{\sqrt{4m_{\pi}^2+q^2}}{q}\,\ln
\frac{\sqrt{4m_{\pi}^2+q^2}+q}{2m_{\pi}}\,,
\nonumber\\&&
A(q)=\frac{1}{2q}\,\arctan\frac{q}{2m_{\pi}}\,.
\end{eqnarray}

However, it does not take into consideration the case $|x|<1$, when 
$t'$ gets closer to $4m_{\pi}^2$. This region, where the naive heavy 
baryon expansion fails, is responsible for the long distance 
behavior of the triangle diagram, as can be seen by its representation
in configuration space,

\begin{eqnarray}
\Gamma(r)&=&\frac{1}{\pi}\int_{4m_{\pi}^2}^{\infty}dt'
\int\frac{d^3q}{(2\pi)^3}\;
e^{-i{\mbox{\boldmath $q$}}\cdot{\mbox{\boldmath $r$}}}\;
\frac{\mbox{Im}\gamma(t')}{t'+q^2}
\nonumber\\&=&
\frac{1}{4\pi^2}\int_{4m_{\pi}^2}^{\infty}dt'\;\frac{e^{-r\sqrt{t'}}}{r}\;
\,\mbox{Im}\gamma(t')\,.
\end{eqnarray}

Therefore it is clear that, in order to have a good asymptotic
description of $\Gamma(r)$, one needs a decent representation for
$\mbox{Im}\gamma(t')$ near $t'=4m_{\pi}^2$. This is only possible if 
one takes the triangle anomalous threshold into account, which cannot be 
provided by current versions of the heavy baryon formalism. 

The potential in configuration space is obtained through a Fourier
transform of the potential in momentum space. There one faces the
problem of non-localities, {\em i.e.}, terms dependent on the variable
${\mbox{\boldmath $z$}}={\mbox{\boldmath $p$}}+{\mbox{\boldmath
$p'$}}$, where ${\mbox{\boldmath $p$}}$ and ${\mbox{\boldmath $p'$}}$
are the initial and final CM momentum of the NN system.  The
relativistic loop integrals, which incorporates the dynamics of the
TPE, also depend on this variable in a non-trivial way.  However,
phenomenologically one learns that such terms are not relevant at low
energies and a Taylor expansion in ${\mbox{\boldmath $z$}}$ is usually
considered\footnote{For instance, in the heavy baryon potential
non-localities show up only at $O(q^4)$ (N${}^3$LO).}. In this case
the Fourier transform can be carried out in an easier way (see, for
instance, Refs.~\cite{Partovi:1969wd,Hoshizaki60}). Generically, in
any spin-isospin channel and up to the considered order in the
RB-expansion the potentials may be written as a function of at most
second order in the total momentum operator.  In this paper we take
the local approximation on the radial part of the potentials and keep
only up to linear terms in the operators.  The remaining
non-localities are fairly small all over the range of interest, which
somehow justifies its exclusion and considerably simplifies our
calculations~\footnote{A rough way of estimating this in coordinate
space is by acting with the operator $\nabla^2 /m_\pi^2$ on the local
function which multiplies the variable ${\mbox{\boldmath $z^2$}}$ and
compare with the local which multiplies the variable ${\mbox{\boldmath
$z^0$}}$. In the region between $0.01 {\rm fm}$ and $2 {\rm fm}$ these
functions behave indeed as $ 1/r^7$ but the ratio between the
non-local and local contribution is at the level of $1-5 \% $.}.

\begin{figure*}[tbc]
\begin{center}
\epsfig{figure=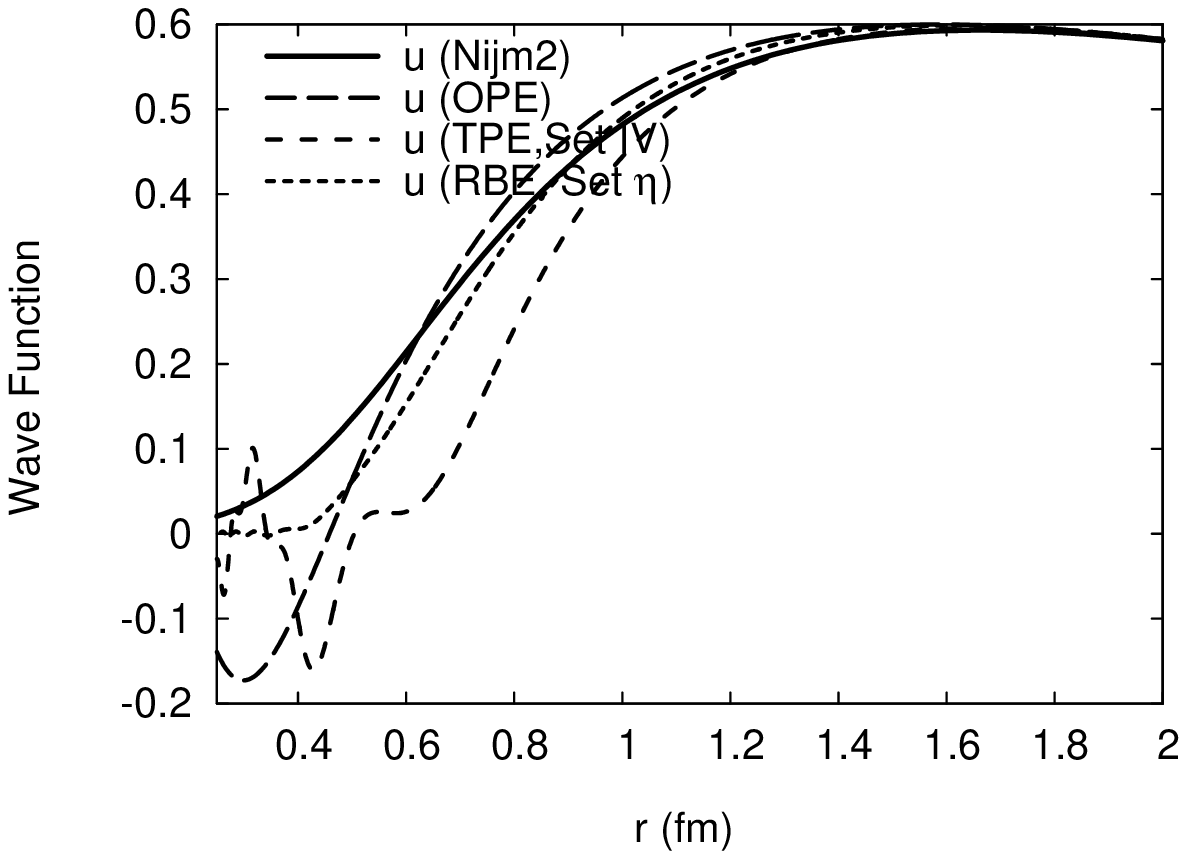,height=6cm,width=8cm}
\epsfig{figure=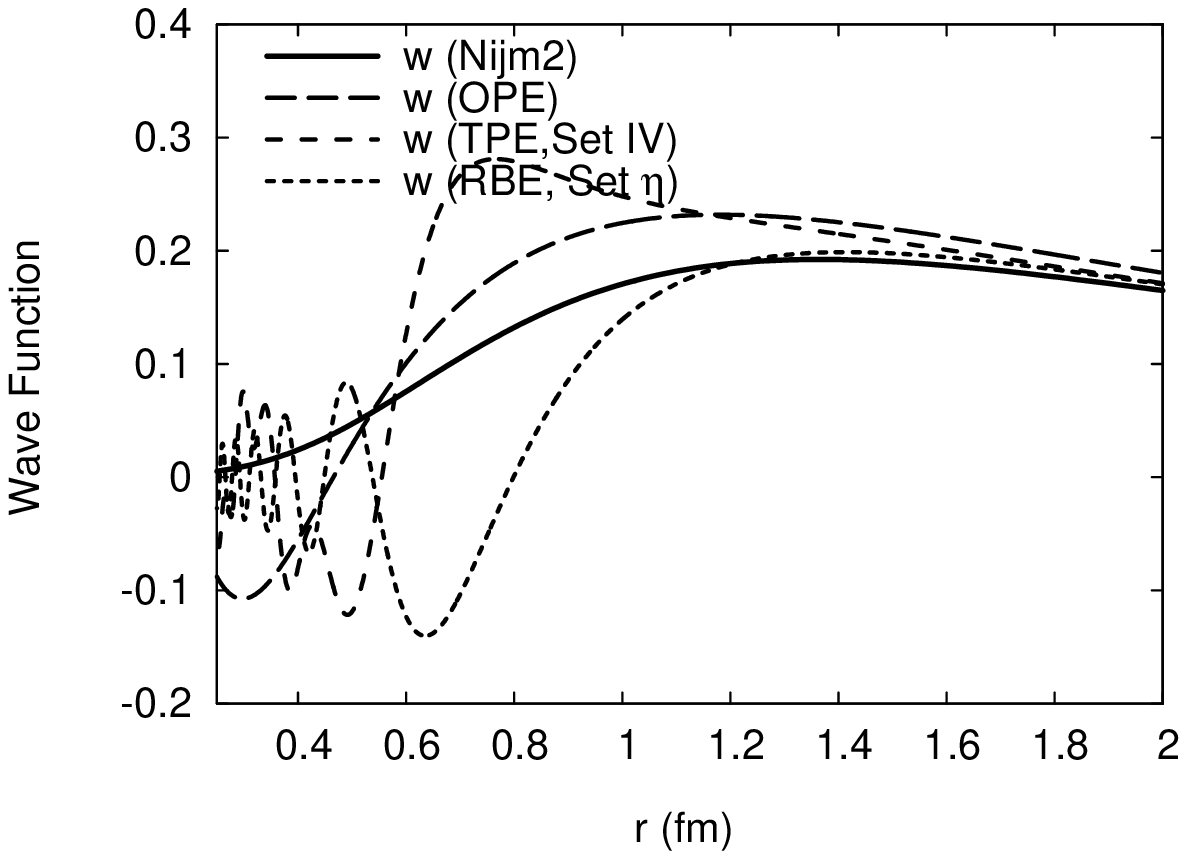,height=6cm,width=8cm}
\end{center}
\caption{RB-TPE deuteron wave functions, u (left) and w (right), as a
function of the distance (in {\rm fm}) compared to the HB-TPE and
Nijmegen II wave functions~\cite{Stoks:1994wp} . The asymptotic
normalization $u \to e^{-\gamma r}$ has been adopted and the
asymptotic D/S ratio is taken $\eta = 0.0256 (4)$ in the TPE case (for
OPE $\eta = 0.026333$).  We use the Set IV of chiral couplings and the
Set $\eta$ (see table~\ref{tab:table1}).}
\label{fig:u+w_TPE}
\end{figure*}

\begin{table*}
\caption{\label{tab:table3} Deuteron properties for the OPE and the
HB- and RB-TPE potentials.  We use the non-relativistic relation $
\gamma= \sqrt{ 2 \mu_{np} B} $ with $B=2.224575(9)$. The errors in the
OPE case are estimated by changing the short distance cutoff in the
range 0.1 - 0.2 fm.  The errors quoted in the HB-TPE reflect the
uncertainty in the non-potential parameters $\gamma$, $\eta$ only. The
errors quoted in the RB-TPE are estimated as in the OPE case, but
changing the cutoff in the range 0.3 - 0.4 fm. The entry Exp. stands for
Experimental and/or recommended values and can be traced from
Ref.~\cite{deSwart:1995ui}.
}
\begin{tabular}{|c|c|c|c|c|c|c|c|c|c|c|}
\hline  Set  & $\gamma ({\rm fm}^{-1})$ & $\eta$ & $A_S ( {\rm fm}^{-1/2}) $
& $r_d ({\rm fm})$ & $Q_d ( {\rm fm}^2) $ & $P_D $ & $\langle r^{-1} \rangle ( {\rm fm}^{-1}) $
& $\langle r^{-2} \rangle ( {\rm fm}^{-2}) $& $\langle r^{-3} \rangle ( {\rm fm}^{-3}) $
\\ \hline\hline 
%
%
{\rm OPE}  & Input & 0.02634 & 0.8681(1) & 1.9351(5) & 0.2762(1)
& 7.88(1)\%  & 0.4861(10) & 0.434(3) & $\infty$ 
\\ \hline
%
%
%
%
%
HB Set IV  & Input & Input & 0.884(4) & 1.967(6) & 0.276(3)
& 8(1)\%  & 0.447(5) & 0.284(8) & 0.276(13) 
\\ \hline
RBE Set IV & Input & 0.03198(3) & 0.8226(5) & 1.8526(10) & 0.3087(2)
& 22.99(13) \%  &  0.5054(15)  & 0.360(4) & 0.333(13)  
\\ \hline 
RBE Set $\eta$ & Input & 0.02566(1) & 0.88426(2) & 1.96776(1) & 0.2749(1)
& 5.59(1) \%  &  0.4438(3)  & 0.2714(7) &0.215(3) 
\\ \hline\hline
NijmII & 0.231605 & 0.02521 & 0.8845(8) & 1.9675 & 0.2707 & 
5.635\%  & 0.4502 & 0.2868 & $\infty$ 
\\
Reid93 & 0.231605 & 0.02514 & 0.8845(8) & 1.9686 & 0.2703 & 
5.699\%   & 0.4515 & 0.2924 & $\infty$ 
\\ \hline \hline 
Exp. &  0.231605 &  0.0256(4)  & 0.8846(9)  & 1.971(6)  &
0.2859(3) & 5.67(4)\%  & - & - & -  
\\ \hline 
\end{tabular}
\end{table*}

\begin{figure}[tbc]
\begin{center}
\epsfig{figure=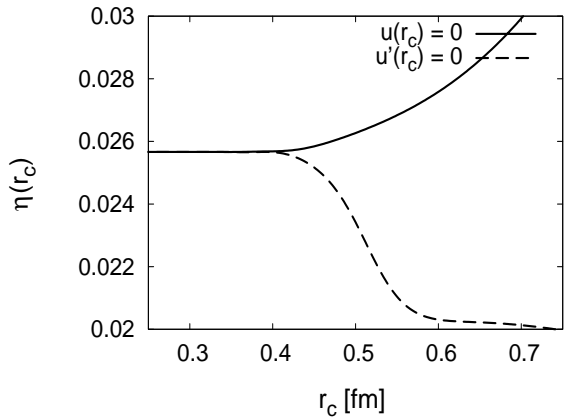,height=6cm,width=8cm}
\end{center}
\caption{Cut-off dependence of the asymptotic D/S ratio of the
deuteron wave function, $\eta$, as a function of the short distance
cut-off $r_c$ (in ${\rm fm}$) for the RB-TPE potential for the
auxiliary boundary conditions $u(r_c)=0$ and $u'(r_c)=0$. We use the
Set $\eta$ of chiral couplings and the Set $\eta$ (see
table~\ref{tab:table1}) designed to reproduce the experimental value
$\eta=0.0256(4)$ in the limit $r_c \to 0$. The convergence towards the
renormalized value is $\eta (r_c) -\eta (0) \sim \exp [ -2/5 (R_+ /
r_c)^\frac52 ] $ up to oscillations (see main text).}
\label{fig:eta-rc}
\end{figure}

\section{The deuteron} 
\label{sec:deuteron} 

In the pn CM system the deuteron wave function is  
\begin{eqnarray}
\Psi (\vec x) &=& \frac1{\sqrt{4\pi}r } \Big[ u(r) \sigma_p \cdot
\sigma_n \nonumber \\ &+& \frac{w(r)}{\sqrt{8}} \left( 3 \sigma_p
\cdot \hat x \, \sigma_n \cdot \hat x - \sigma_p \cdot \sigma_n \right)
\Big] \chi_{pn}^{s m_s}
\end{eqnarray} 
with the total spin $s=1$ and $m_s=0,\pm 1$ and $\sigma_p$ and
$\sigma_n$ the Pauli matrices for the proton and the neutron
respectively. The functions $u(r)$ and $w(r)$ are the reduced S- and
D-wave components of the relative wave function respectively. They 
satisfy the coupled set of equations in the $^3S_1 - ^3D_1 $ channel
\begin{eqnarray}
-u '' (r) + U_{^3S_1} (r) u (r) + U_{E_1} (r) w (r) &=& -\gamma^2 u
 (r) \, ,\nonumber \\ -w '' (r) + U_{E_1} (r) u (r) + \left[U_{^3D_1}
 (r) + \frac{6}{r^2} \right] w (r) &=& -\gamma^2 w (r) \, , \nonumber
 \\
\label{eq:sch_coupled} 
\end{eqnarray}
with $ U_{^3S_1} (r)$, $U_{E_1} (r) $ and $U_{^3D_1} (r)$ the
corresponding matrix elements of the coupled channel potential.
We solve Eq.~(\ref{eq:sch_coupled}) together with the asymptotic
condition at infinity
\begin{eqnarray}
u (r) &\to & A_S e^{-\gamma r} \, , \nonumber \\ w (r) & \to & A_D
e^{-\gamma r} \left( 1 + \frac{3}{\gamma r} + \frac{3}{(\gamma r)^2}
\right) \, ,
\label{eq:bcinfty_coupled} 
\end{eqnarray}
where $ \gamma = \sqrt{M B} $ is the deuteron wave number ($B$ is the
deuteron binding energy), $A_S$ is the s-wave normalization factor
determined from the condition
\begin{eqnarray}
\int_0^\infty dr \left[ u(r)^2 + w(r)^2 \right] = 1  \, ,  
\label{eq:normalization} 
\end{eqnarray}
and the asymptotic D/S ratio parameter is defined by $\eta=A_D/A_S$.
As we have mentioned already the RB-TPE potential displays a
relativistic $1/r^7$ Van der Waals singularity at the origin. Thus,
the discussion on whether or not the deuteron parameters $\gamma$ and
$\eta$ can be fixed independently on the potential depends on the
short distance behaviour of the eigenvalues of the coupled channel
potential matrix.  As discussed in
Ref.~\cite{PavonValderrama:2005gu,Valderrama:2005wv,PavonValderrama:2005uj,Erratum}
also for the bound state case the number of independent parameters
coincides with the number of negative (attractive) eigenpotentials at
short distances. In the RB-TPE potential we are using
here~\cite{Higa:2003jk,Higa:2003sz, Higa:2004cr} we have only one
independent parameter (see table~\ref{tab:table2}) which we take to be
$\gamma$ or, equivalently, the deuteron binding energy. With such a
choice $\eta$ becomes a prediction in contrast to the HB-TPE where
$\eta$ is an input. The outgoing deuteron wave functions are
depicted in Fig.~\ref{fig:u+w_TPE} for the RB-TPE potential and
compared to the HB-TPE one.

Let us analyze in more detail the cut-off dependence of observables in
the present RB-TPE potential. Given the fact that in the $^3S_1-^3D_1$
coupled channel we have one attractive and one repulsive
eigenpotential at short distances we may borrow from the previous
discussion on OPE~\cite{PavonValderrama:2005gu} where we refer for
further details.  The practical way of introducing in this case a
short distance cut-off $r_c$ which selects the regular solution at the
origin is by appropriately choosing the auxiliary boundary  condition at the
point $r=r_c$ among many possible choices compatible with
self-adjointness~\cite{PavonValderrama:2005gu}. The precise choice may
provide smoother limits and hence better convergence properties in the
pre-asymptotic region. Actually, as we show in
Appendix~\ref{finite-cut-off}, we can estimate the size of the finite
cut-off corrections in deuteron observables and hence their
convergence rate towards the corresponding renormalized values. The
result is that up to some oscillations the convergence towards the
renormalized value is exponential as $r_c \to 0$, i.e.  the
convergence towards the renormalized value is $\eta (r_c) -\eta (0)
\sim \exp [ -4/5 (R_+ / r_c)^\frac52 ] $ for the RB-TPE potential (
$\sim 1/r^7$ ) as compared to $\eta (r_c) -\eta (0) \sim \exp [ -2
(R_+ / r_c)^\frac12 ] $ for the OPE potential ( $\sim 1/r^3$). Here $
R_+$ is a characteristic short distance scale of the corresponding
repulsive eigenpotentials.  The analysis of
Appendix~\ref{finite-cut-off} also shows that, generally, the more
singular the potential the faster the convergence. 

For illustration purposes we show in Fig.~\ref{fig:eta-rc} the
calculated asymptotic D/S ratio of the deuteron wave function, $\eta$,
as a function of the short distance cut-off $r_c$ (in ${\rm fm}$) for
the RB-TPE potential when the auxiliary boundary conditions $u
(r_c)=0$ and $u'(r_c)=0$ are considered. Note the clear and coincident
plateau below $r_c=0.4 {\rm fm}$ for both boundary conditions
following a relatively rapid variation above this
region~\footnote{Other auxiliary boundary conditions such as
$w(r_c)=0$ or $w'(r_c)=0$ not shown in the figure make the
oscillations deduced in Appendix~\ref{finite-cut-off} more visible,
but the plateau region takes place also below $r_c=0.4 {\rm fm}$ and
yields an identical numerical result as with $u(r_c)=0$ and
$u'(r_c)=0$.}. As mentioned above, this is a typical feature of a
coupled channel singular potential with one attractive and one
repulsive eigenpotential which extends to all other deuteron
properties.  Actually, the situation strongly resembles the previously
studied OPE potential which has a softer $1/r^3$ singularity at the
origin~\cite{PavonValderrama:2005gu}; the main difference is that for
OPE stability takes place at a shorter scale, $r_c=0.2 {\rm fm}$, than
in RB-TPE potential. In the present context it is also helpful to
remind that a short distance cut-off radius and a sharp momentum
cut-off are inverse proportional to each other, $r_c = \pi /( 2
\Lambda) $~\cite{PavonValderrama:2004td} (the numerical coefficient
depends on the particular regularization). Thus plotting observables
as a function of $r_c$ enhances the finite cut-off changes while long
plateaus could be observed instead as functions of
$\Lambda$. Actually, halving the short distance cut-off corresponds to
doubling the momentum space cut-off. For instance, in RB (Set-$\eta$)
the range $r_c=0.3-0.4 {\rm fm}$ corresponds to the range
$\Lambda=780-1030 {\rm MeV}$ where observables such as $\eta$ change
less than $0.01 \%$.  Pinning down this error bar would be harder if
the sharp or other momentum space cut-off was used.

Our results for several deuteron properties are shown in
table~\ref{tab:table3} and compared to the corresponding HB-TPE
considered in our previous
work~\cite{PavonValderrama:2005gu,Valderrama:2005wv,PavonValderrama:2005uj,Erratum}. Some
remarks concerning the errors quoted in table~\ref{tab:table3} are in
order. We provide the largest source of error in the calculation which
we are able to quantify. Since we aim at renormalized results, we stop
whenever the change of the cut-off causes no significant variation
within a prescribed accuracy, which we take to be about an order of
magnitude higher than the typical experimental or recommended value
uncertainty. Thus, the cut-off range is not necessarily the same in
all cases.  In general terms, the more singular the potential at short
distance, the faster the convergence of the result towards the
renormalized limit (see e.g. Appendix~\ref{finite-cut-off}). Thus, the
toughest case is OPE, where we only have $1/r^3$
singularity. Convergence in this case is the slowest, therefore
shorter cut-offs $r_c=0.1-0.2 {\rm fm}$ are needed.

\begin{figure*}[tbc]
\begin{center}
\epsfig{figure=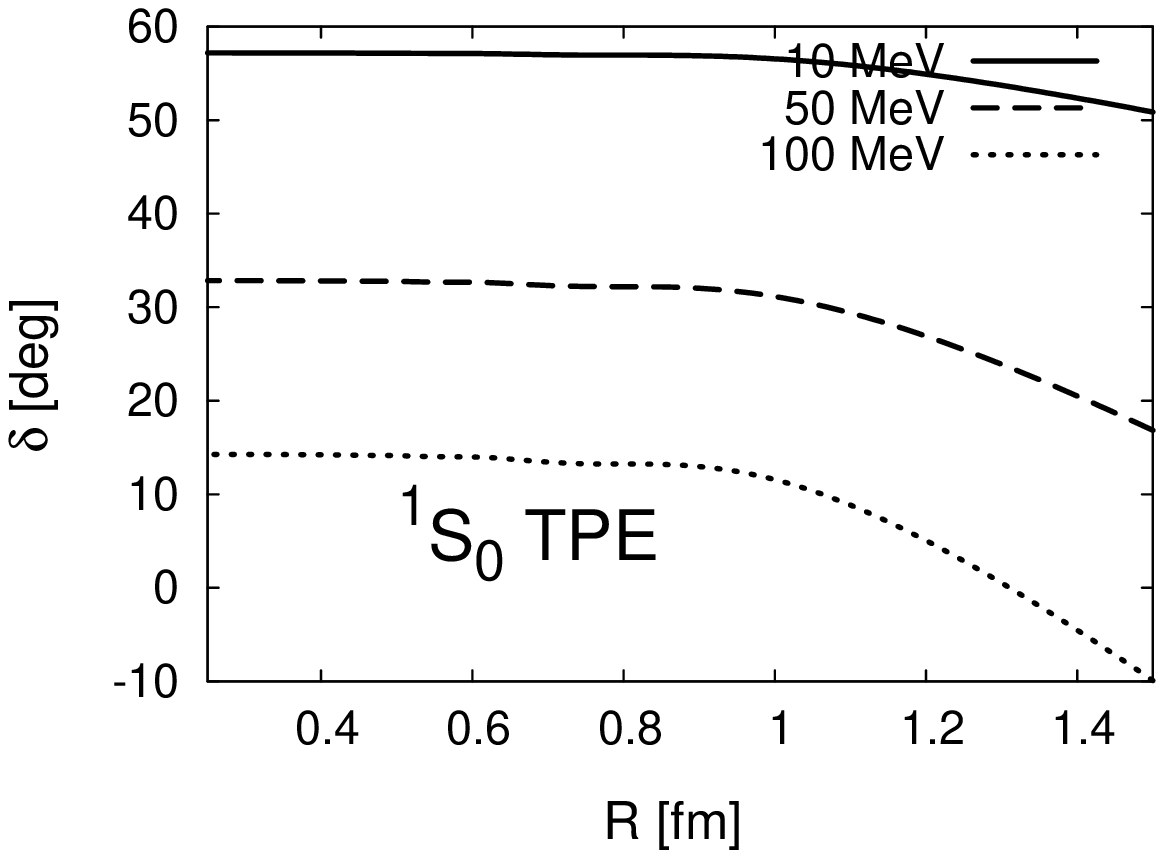, height=4cm, width=5cm}
\epsfig{figure=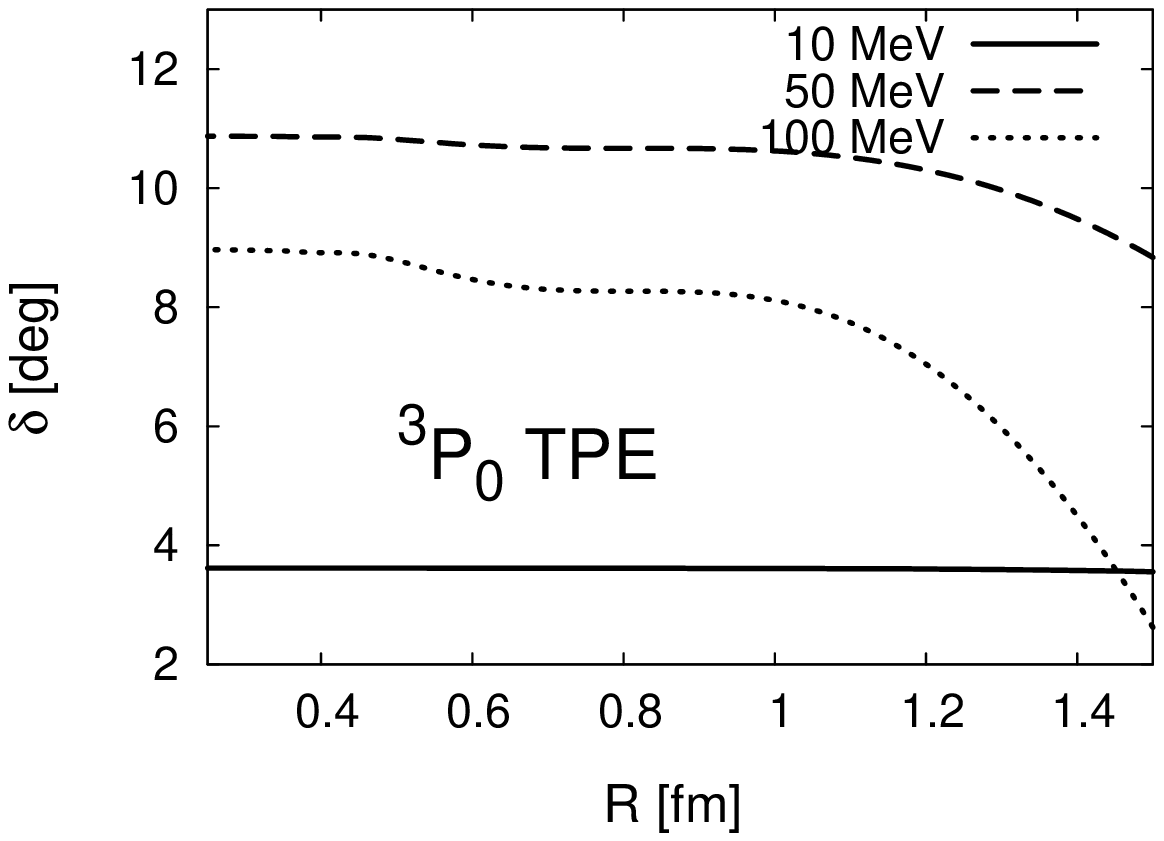, height=4cm, width=5cm}
\epsfig{figure=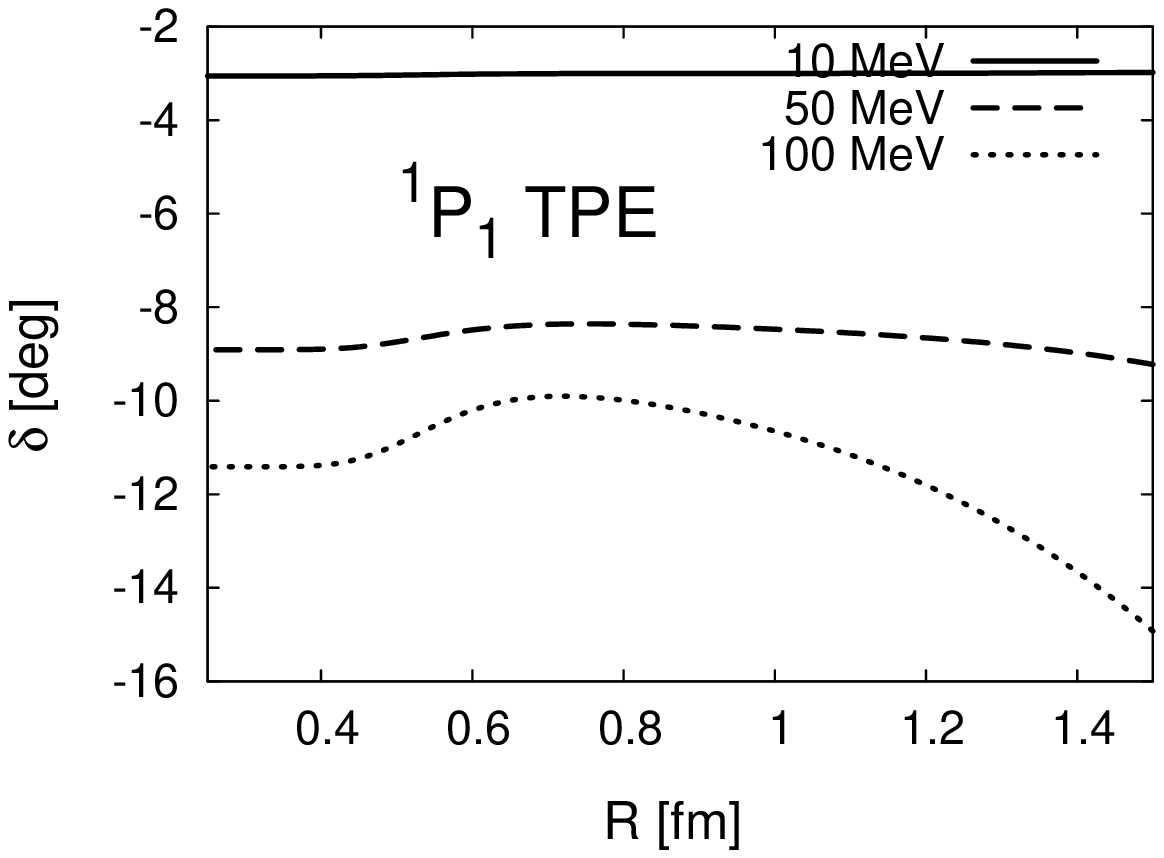, height=4cm, width=5cm}
\end{center}
\caption{Relativistic Baryon expansion (RBE) np phase shifts in the
$^1S_0$, $^3P_0$ and $^1P_1$ channels as a
function the short distance cut-off radius $r_c$ for the fixed
laboratory energies $T_L=10,50,100 {\rm MeV}$.}
\label{fig:ps-running}
\end{figure*}

The value we obtain for $\eta$ for the parameter Sets of
table~\ref{tab:table1} is slightly different from the experimental
one, making the comparison with the HB-TPE
case~\cite{Valderrama:2005wv}, where $\eta$ was a free parameter (two
attractive short distance eigenpotentials), a bit misleading. In order
to obtain an accurate value of $\eta$ it was necessary to readjust
the low energy parameters $c_3$ and $c_4$ to the values $c_3=-3.8 {\rm
GeV}^{-1}$ and $c_4 = 4.5 {\rm GeV}^{-1}$, indeed, very similar to the
values proposed by other
authors~\cite{Buettiker:1999ap,Rentmeester:1999vw,
Epelbaum:2003xx,Entem:2003ft} (see table~\ref{tab:table1}). Actually,
once we have reproduced $\eta$ we see a general and slight improvement
in accuracy when going from the HB-TPE (where $\eta$ is a free
parameter) of our previous work~\cite{Valderrama:2005wv} to the
present RB-TPE calculation (where $\eta$ is predicted). Basically,
RB-TPE produces a sharp prediction for $\eta$ (with eventually no
errors), whereas HB-TPE does not predict $\eta$, so its $1\%$
experimental uncertainty propagates to other deuteron observables at
about a similar $1\%$ level, which is still comparable or larger than
the error in the quoted experimental or recommended values. The
conclusion in Ref.~\cite{Valderrama:2005wv} was that agreement was
partly achieved because of this fuzziness in the theoretical
predictions. Of course, one should not overstress the possible
accuracy of the present results as regards the systematic errors; the main
point of our calculation is to provide the general picture when more
complete asymptotic TPE effects are correctly taken into
account.

\begin{figure*}[tbc]
\begin{center}
\epsfig{figure=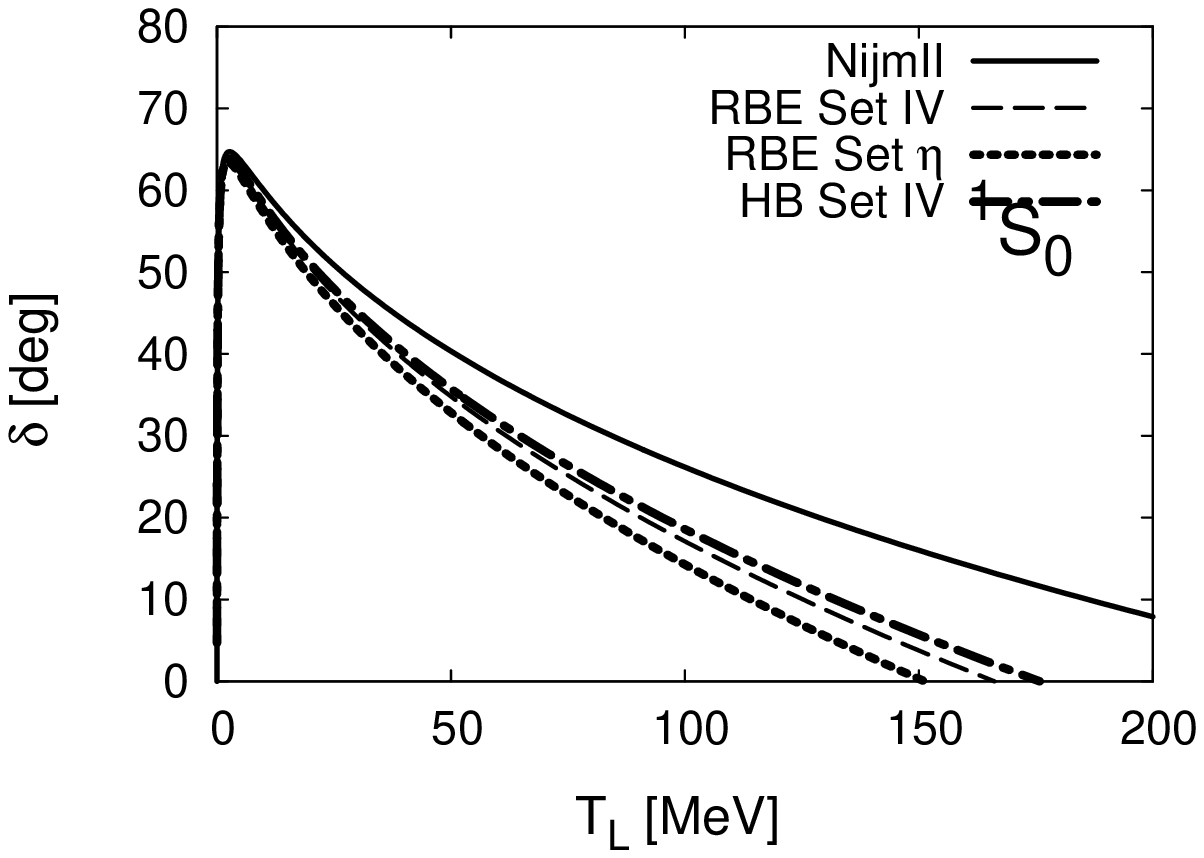, height=4cm, width=5cm}
\epsfig{figure=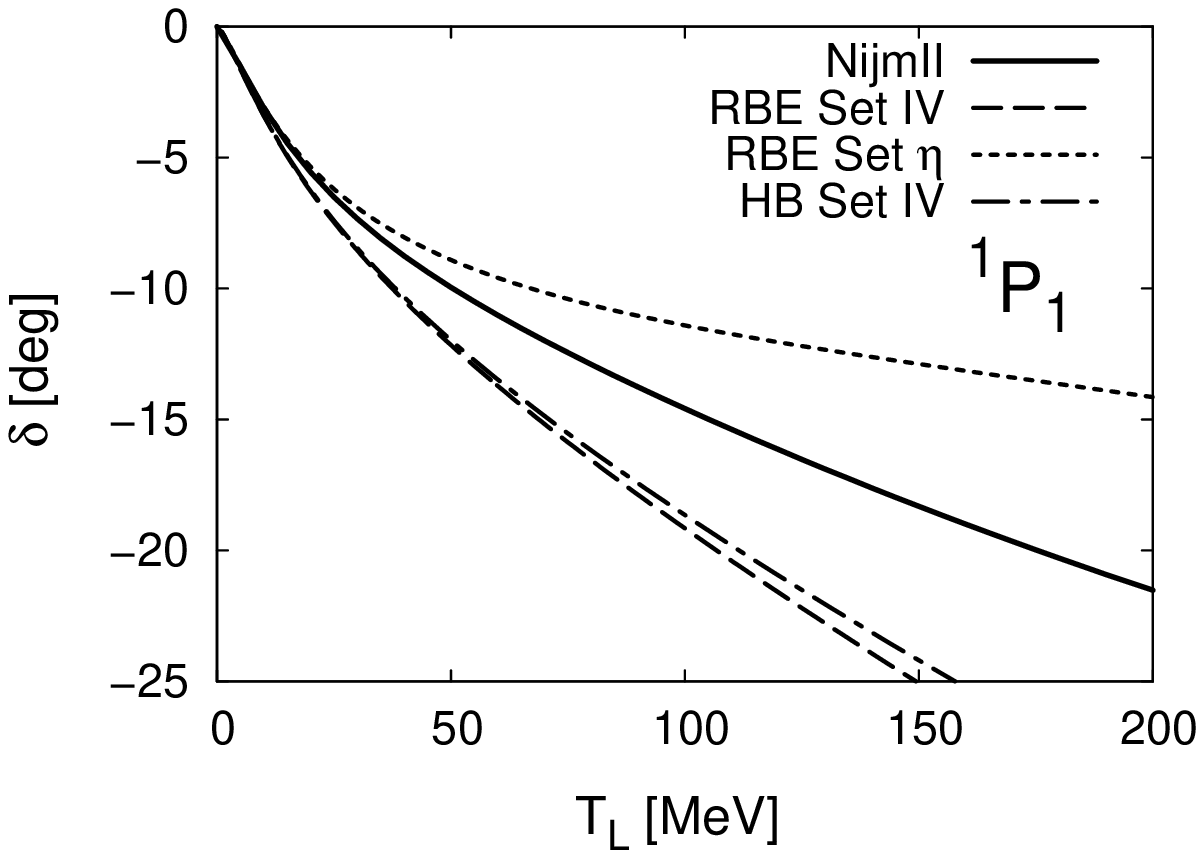, height=4cm, width=5cm}
\epsfig{figure=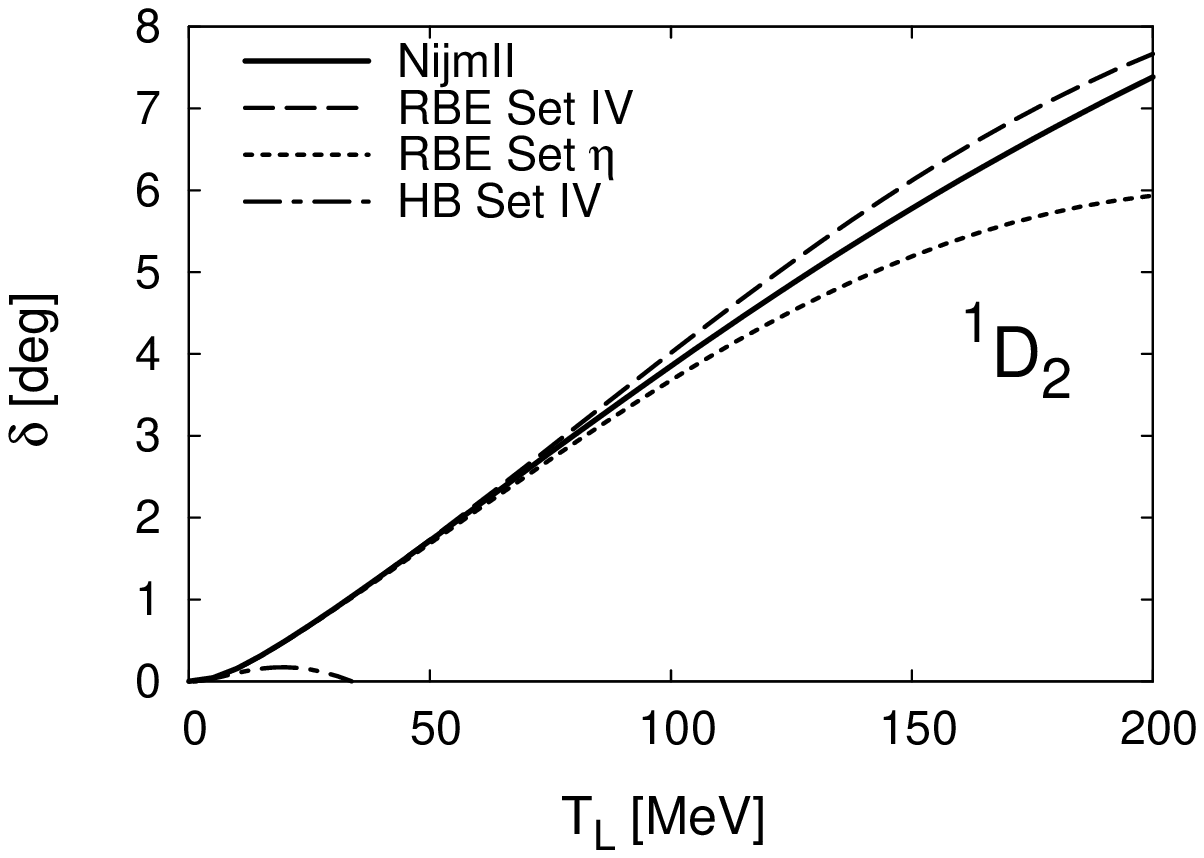, height=4cm, width=5cm}\\ 
\epsfig{figure=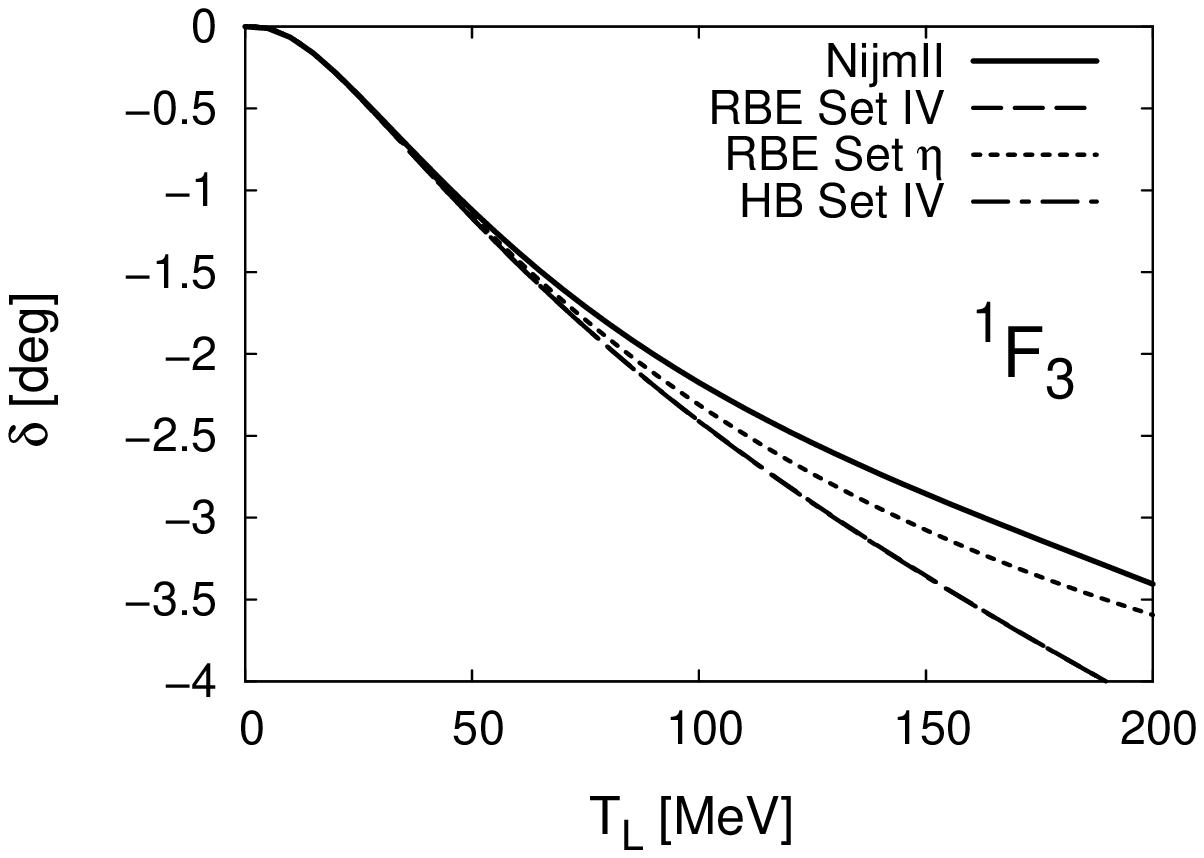, height=4cm, width=5cm}
\epsfig{figure=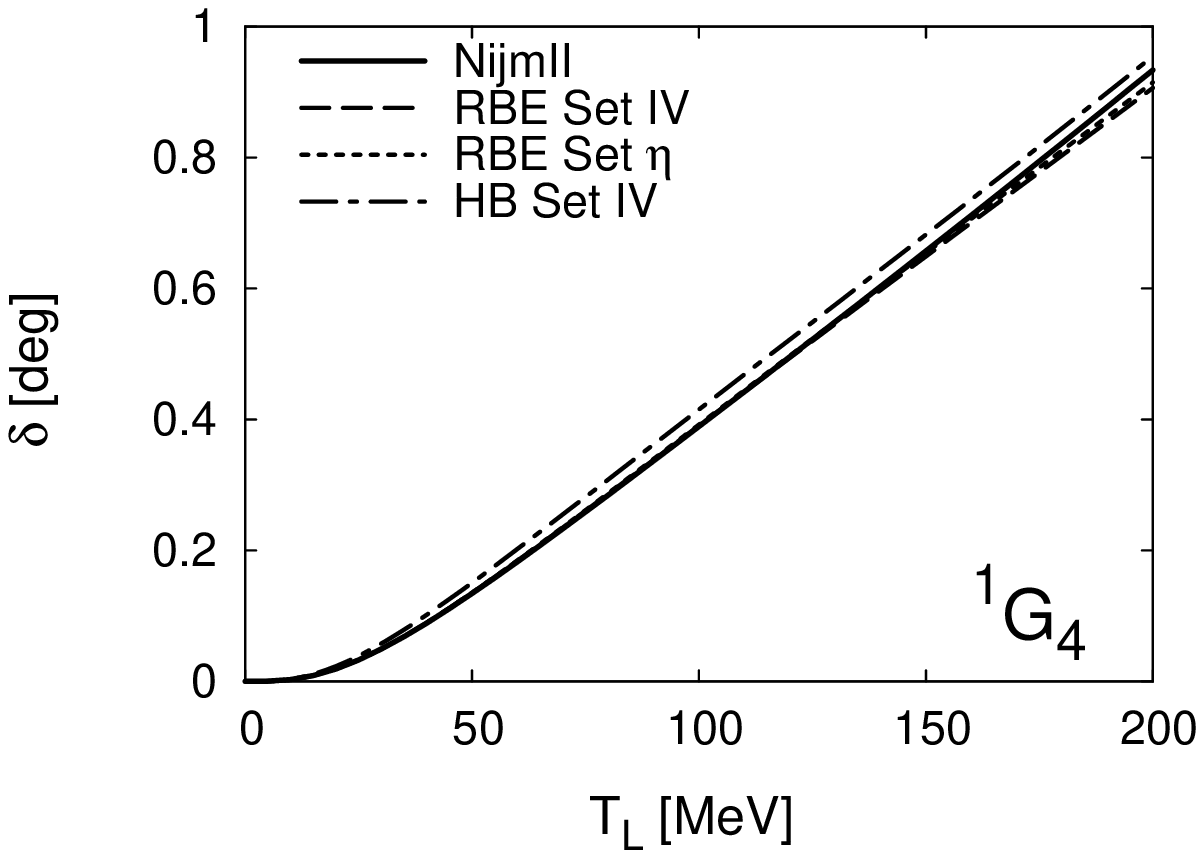, height=4cm, width=5cm}
\epsfig{figure=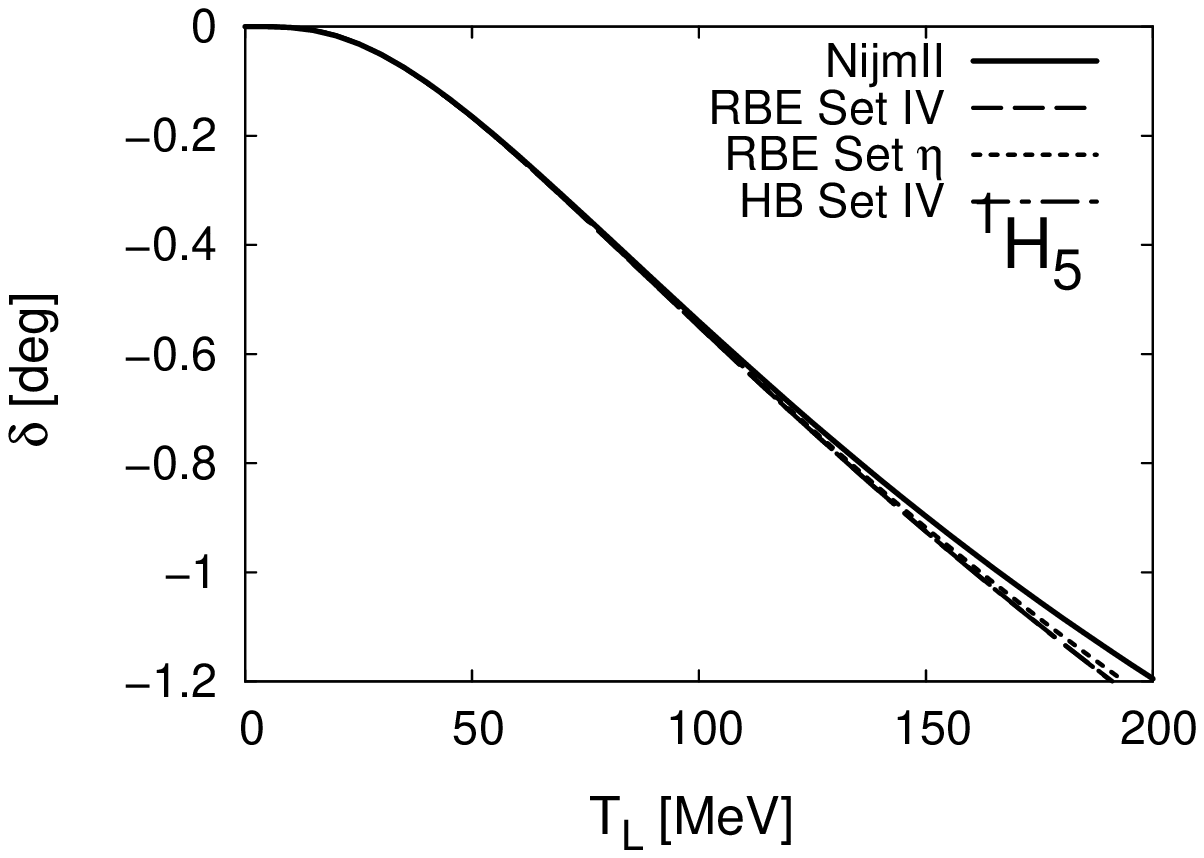, height=4cm, width=5cm}\\ 
\epsfig{figure=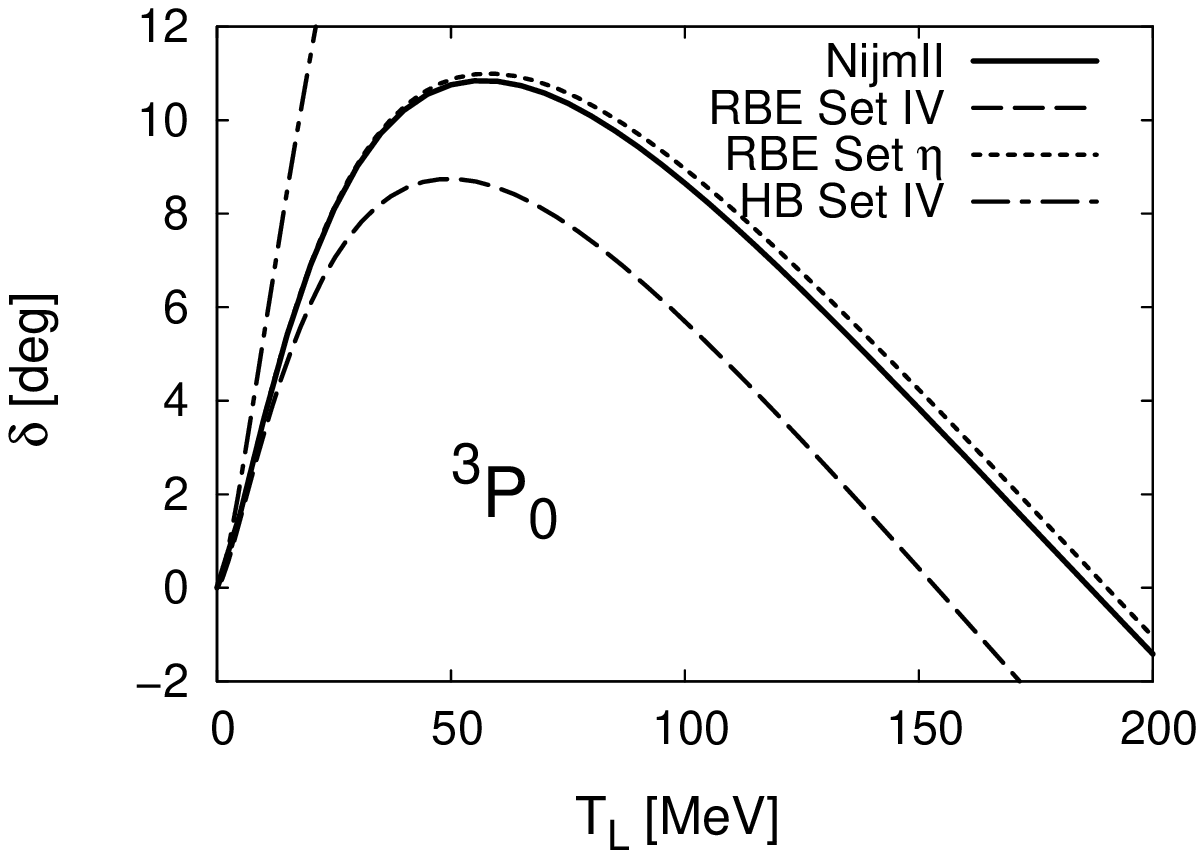, height=4cm, width=5cm} 
\epsfig{figure=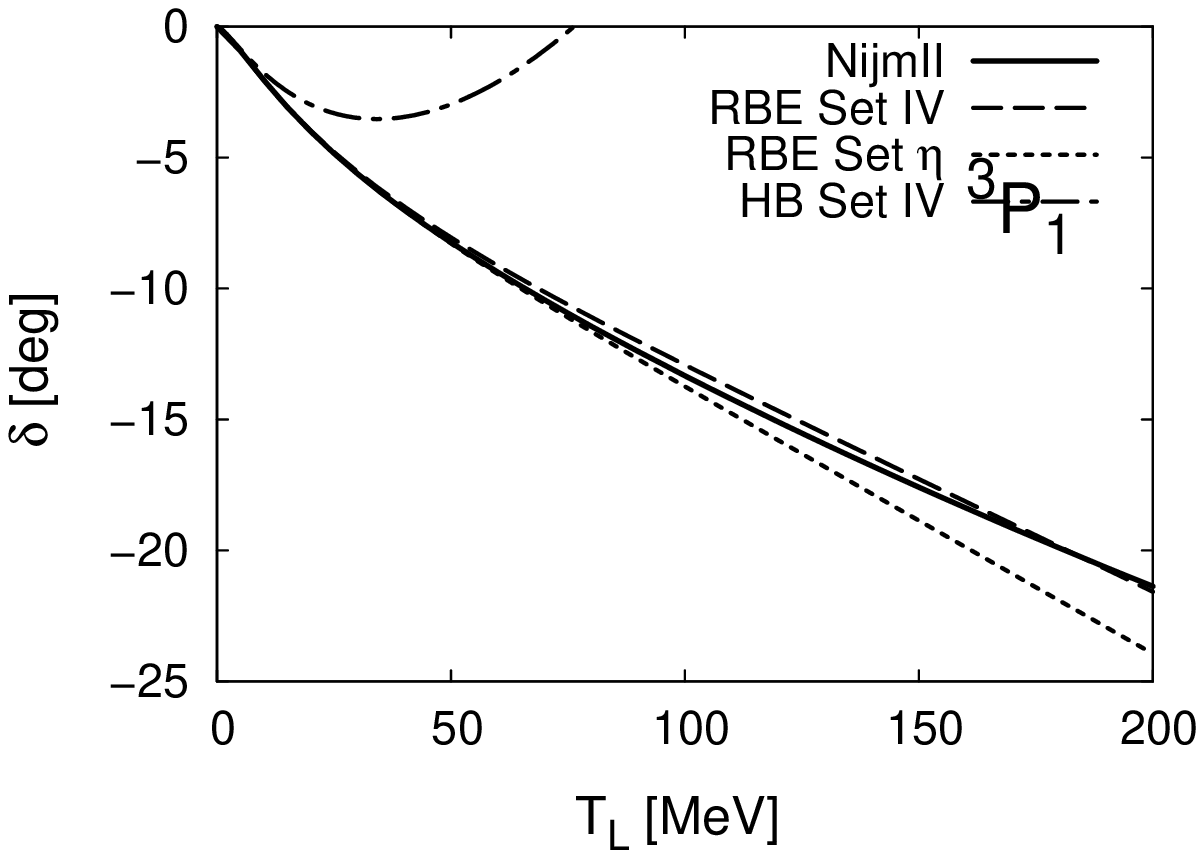, height=4cm, width=5cm}
\epsfig{figure=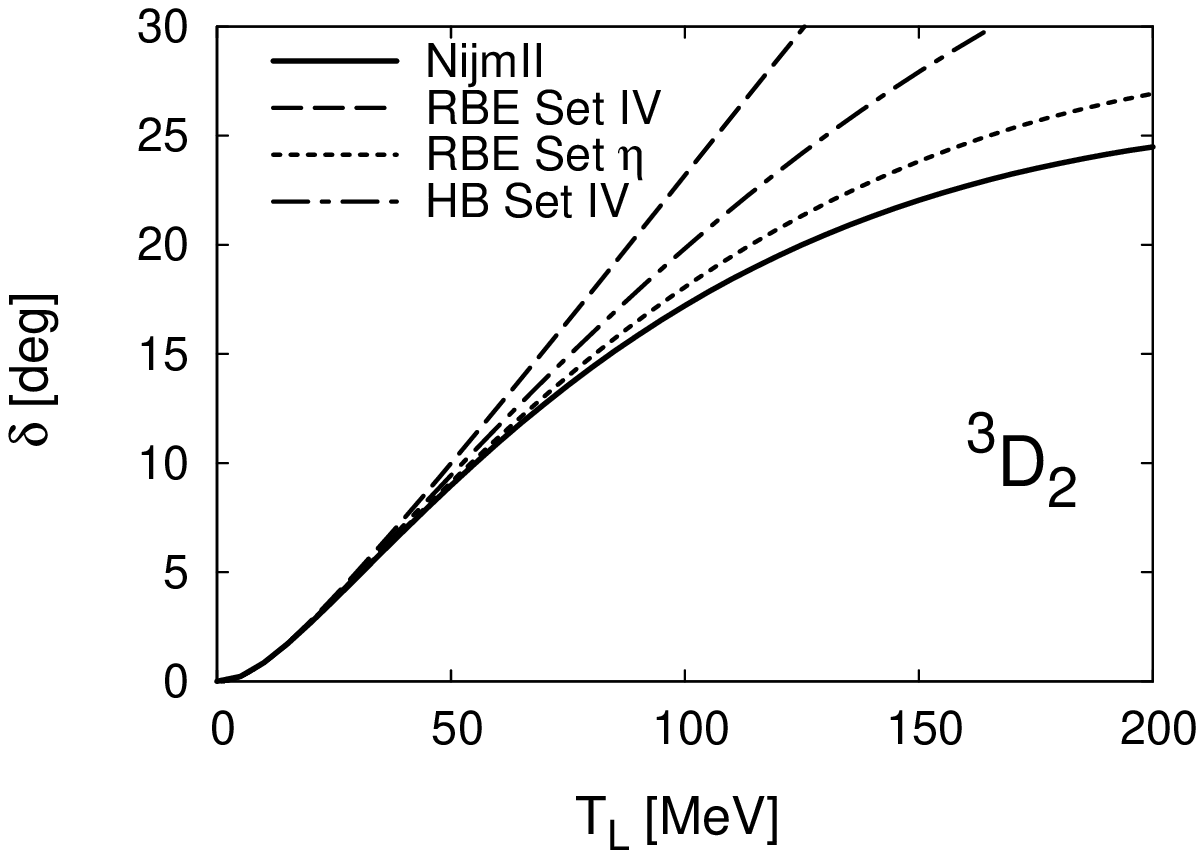, height=4cm, width=5cm}\\
\epsfig{figure=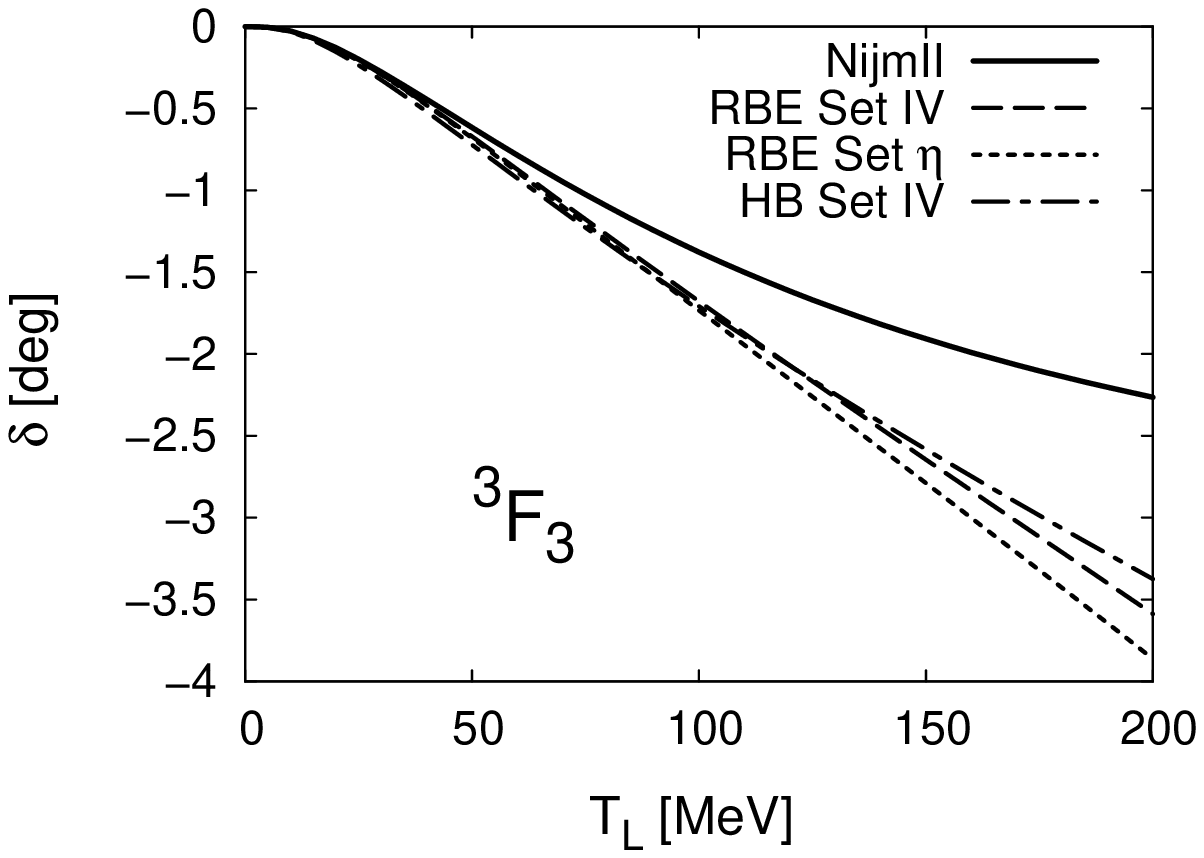, height=4cm, width=5cm}
\epsfig{figure=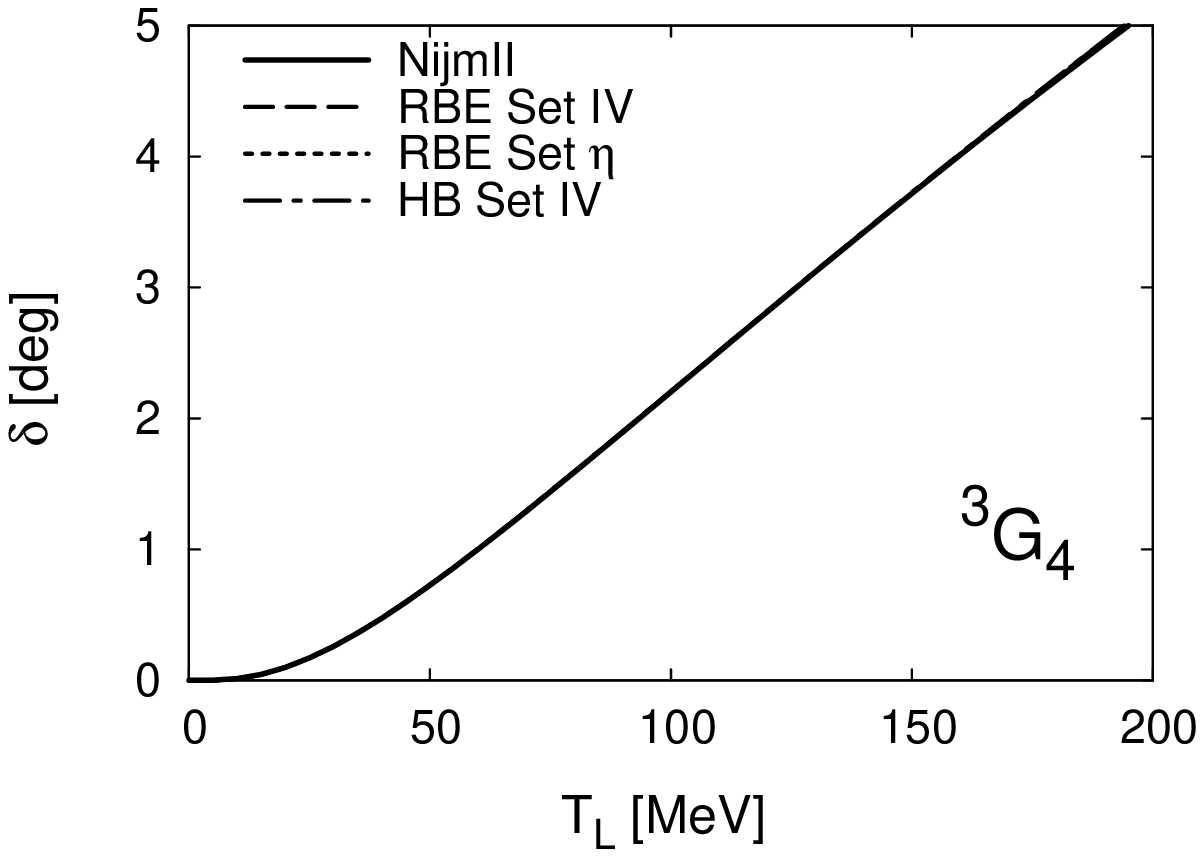, height=4cm, width=5cm}
\epsfig{figure=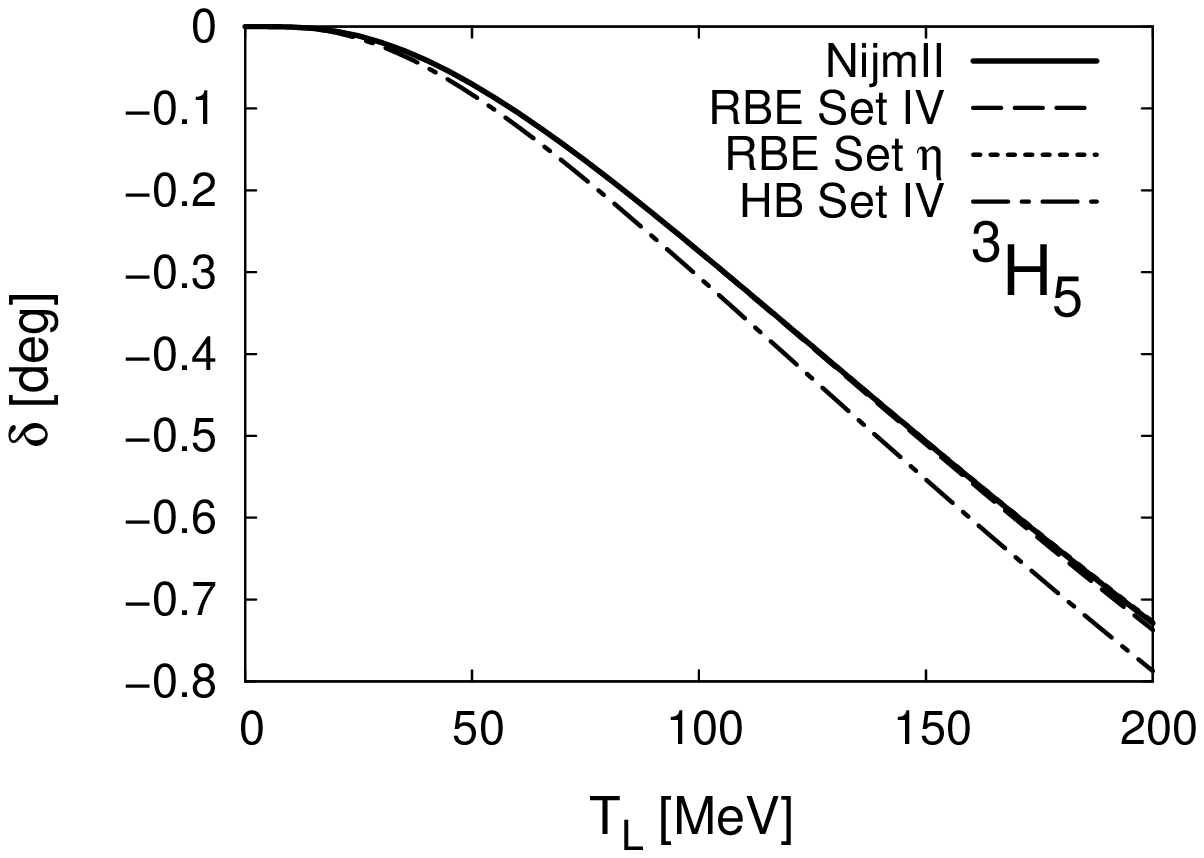, height=4cm, width=5cm}
\end{center}
\caption{np (SYM-nuclear bar) Spin Singlet and Uncoupled Spin Triplet phase
shifts for the total angular momentum $j=0,1,2,3,4,5$ for relativistic
baryon expansion (RBE) and for the heavy baryon expansion (HBE) as a
function of as a function of the LAB energy compared to the Nijmegen
partial wave analysis~\cite{Stoks:1993tb,Stoks:1994wp}.}
\label{fig:cmp-uncoupled}
\end{figure*}

\begin{figure*}[tbc]
\begin{center}
\epsfig{figure=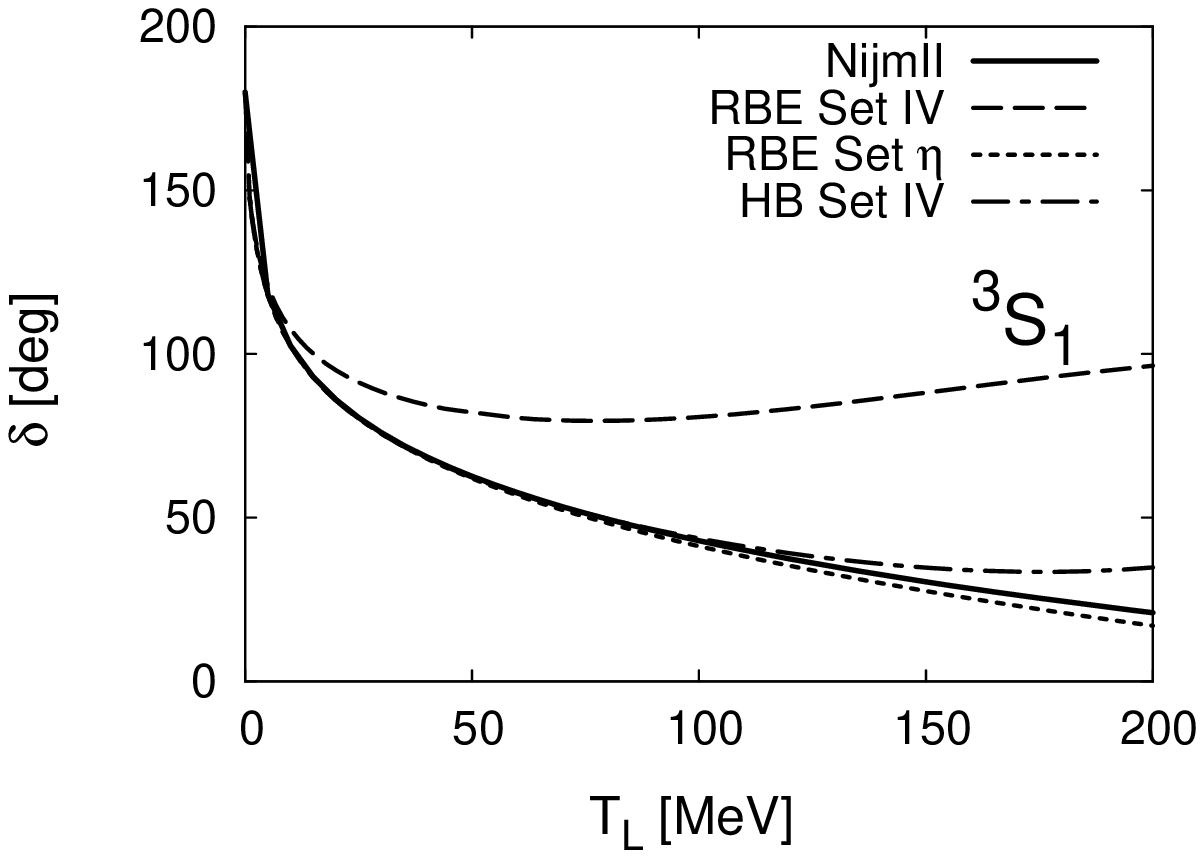, height=4cm, width=5cm}
\epsfig{figure=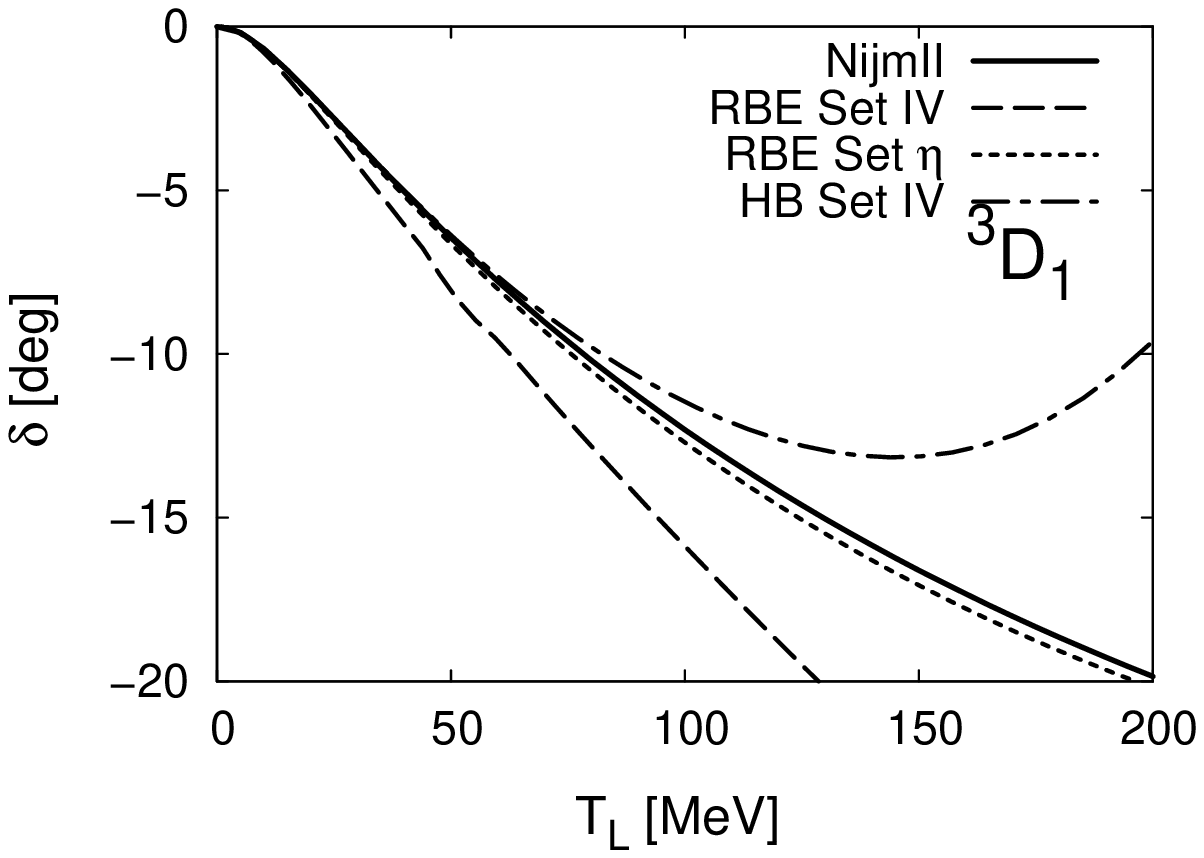, height=4cm, width=5cm}
\epsfig{figure=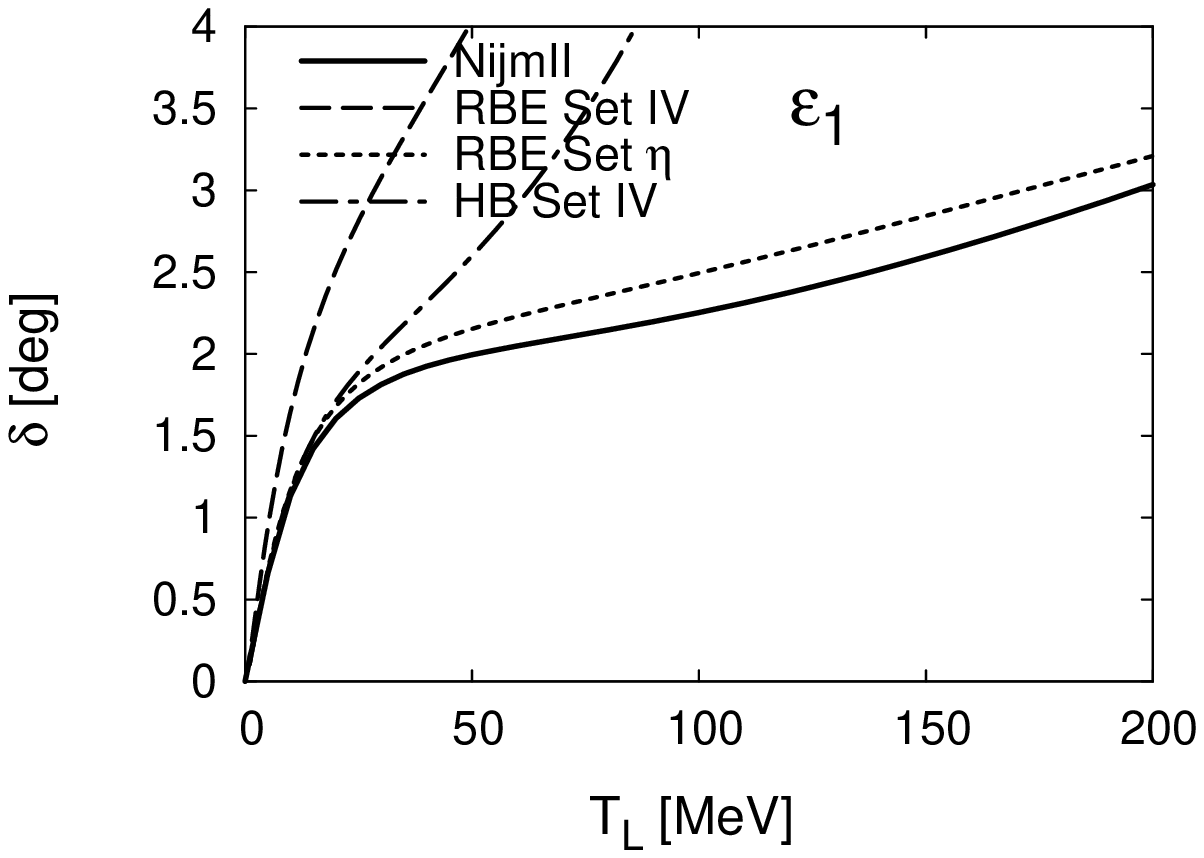, height=4cm, width=5cm} \\
\epsfig{figure=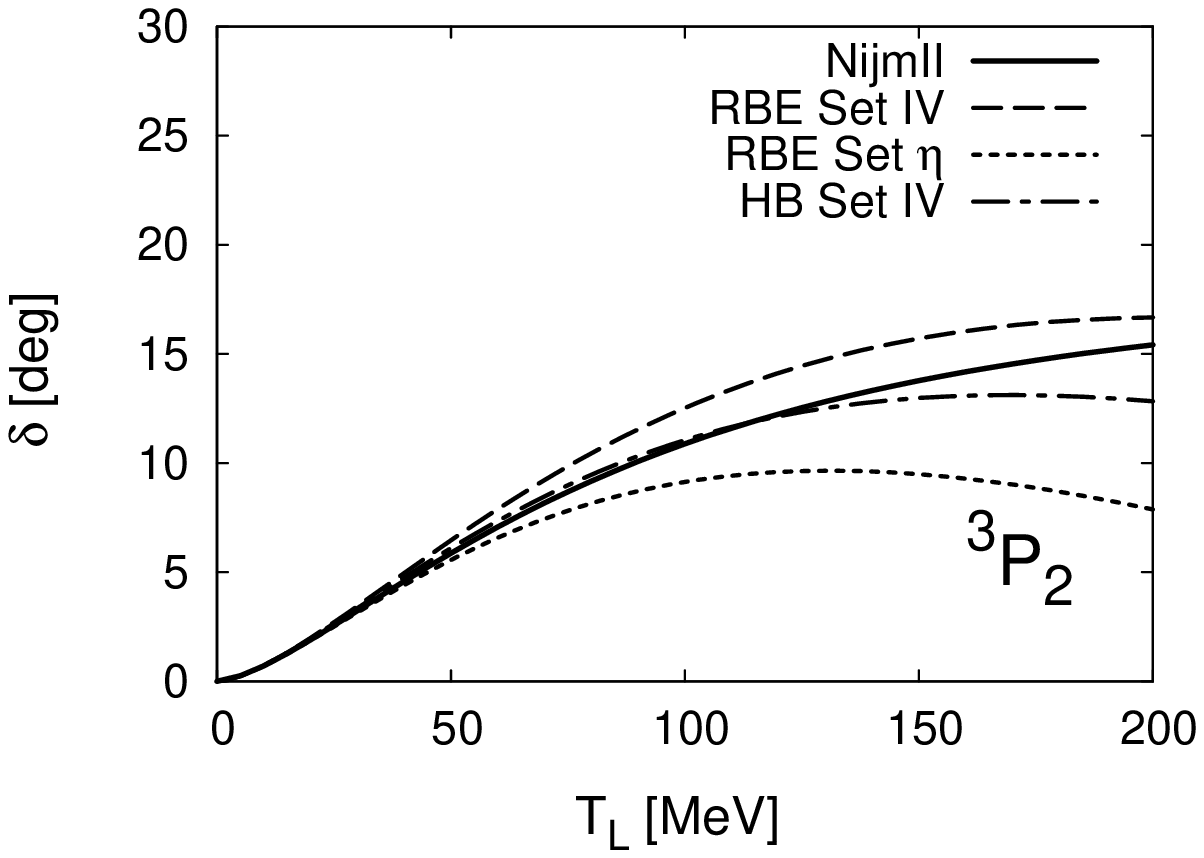, height=4cm, width=5cm}
\epsfig{figure=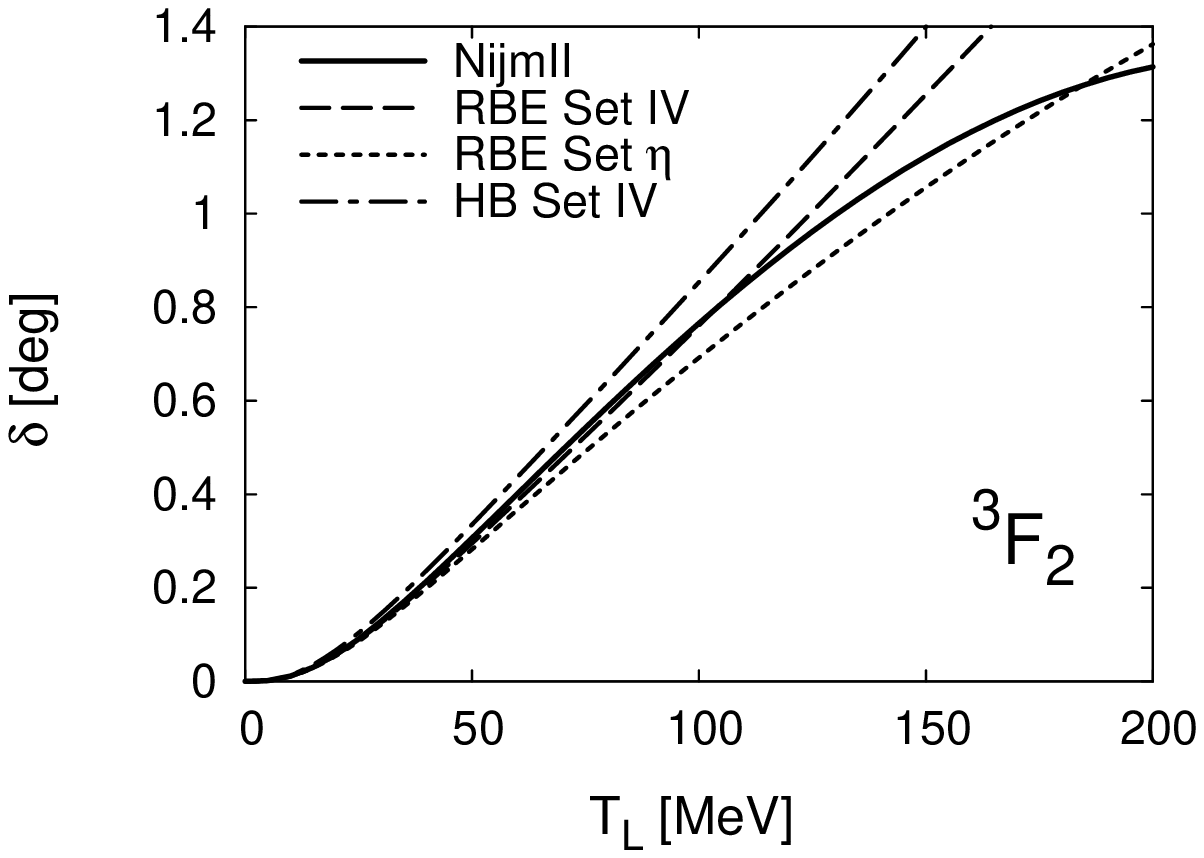, height=4cm, width=5cm}
\epsfig{figure=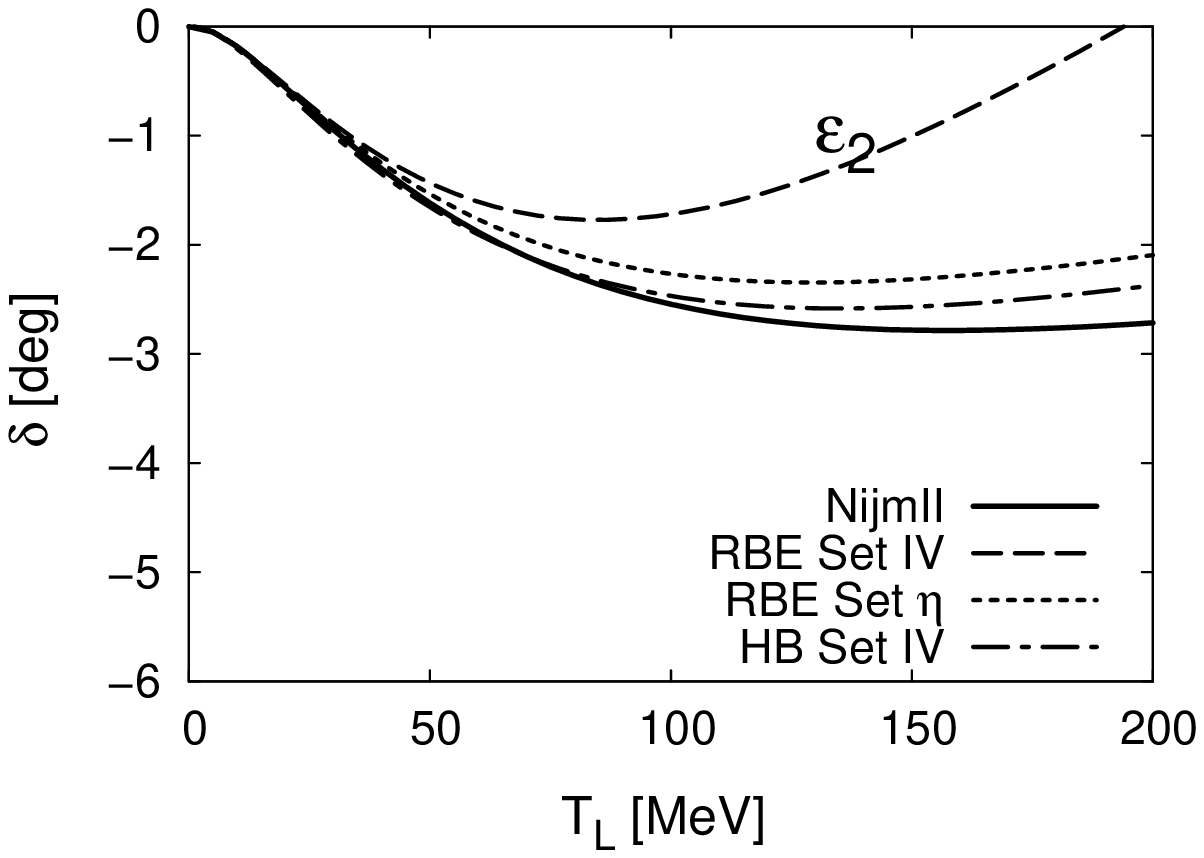, height=4cm, width=5cm} \\ 
\epsfig{figure=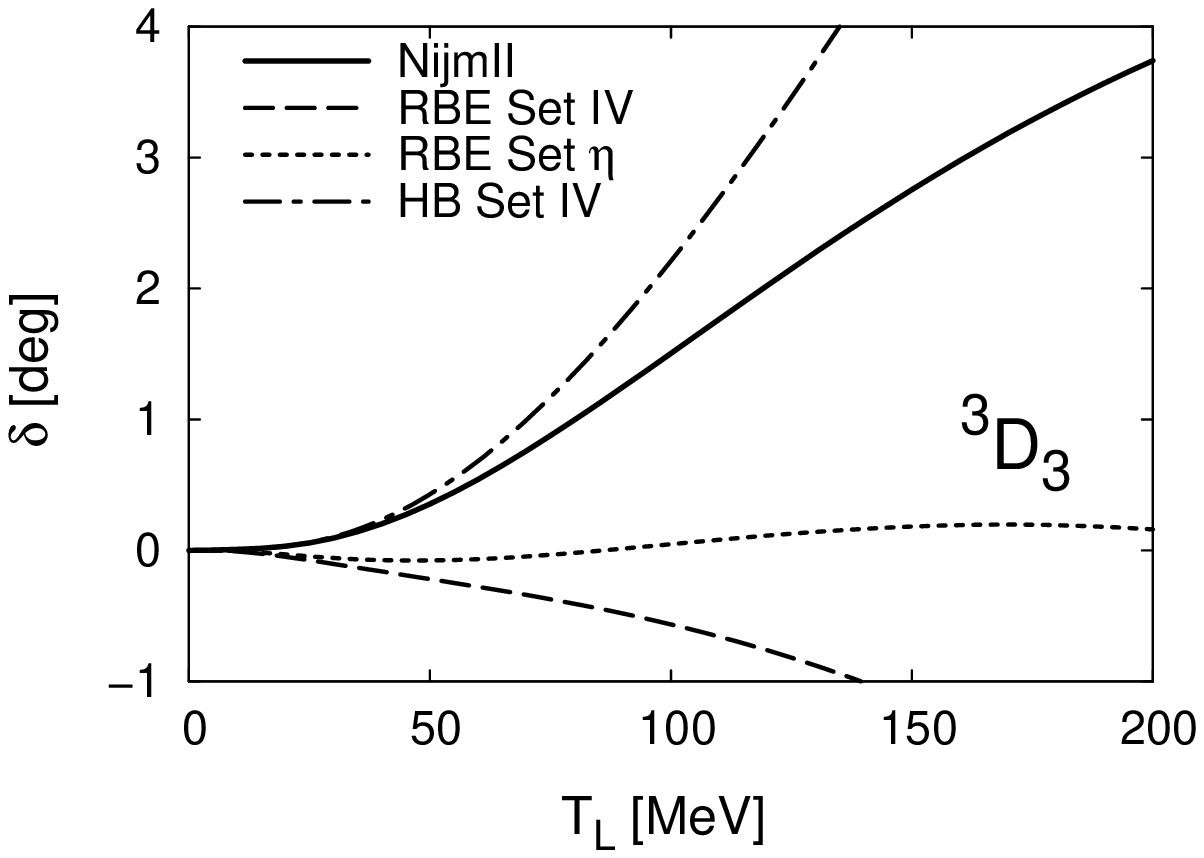, height=4cm, width=5cm}
\epsfig{figure=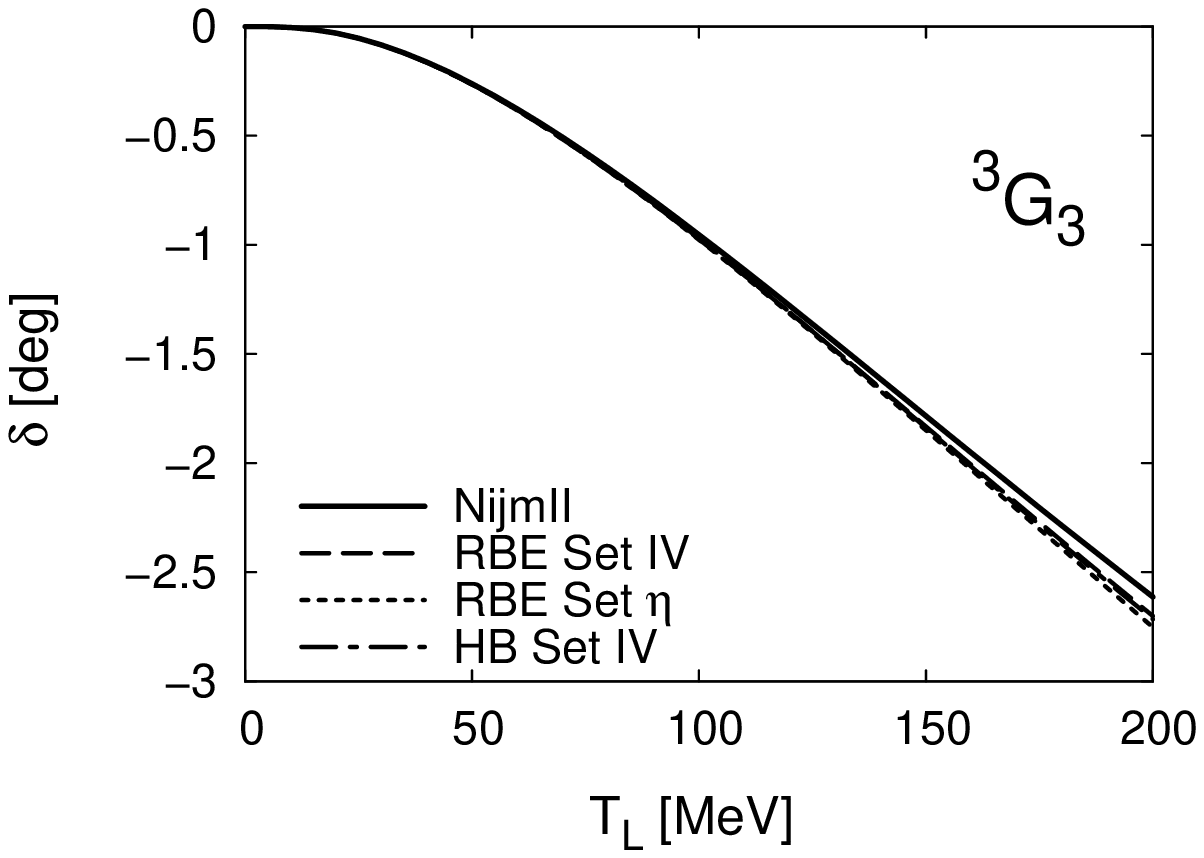, height=4cm, width=5cm}
\epsfig{figure=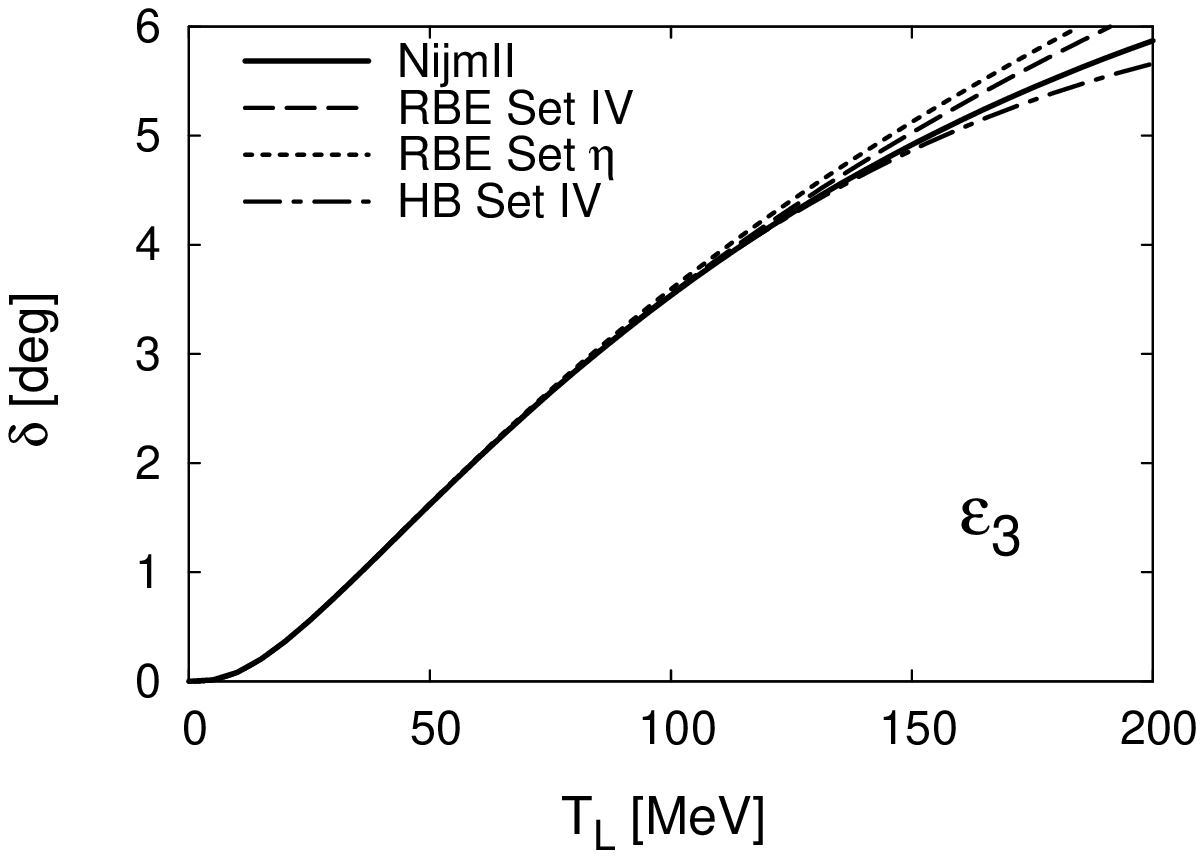, height=4cm, width=5cm} \\ 
\epsfig{figure=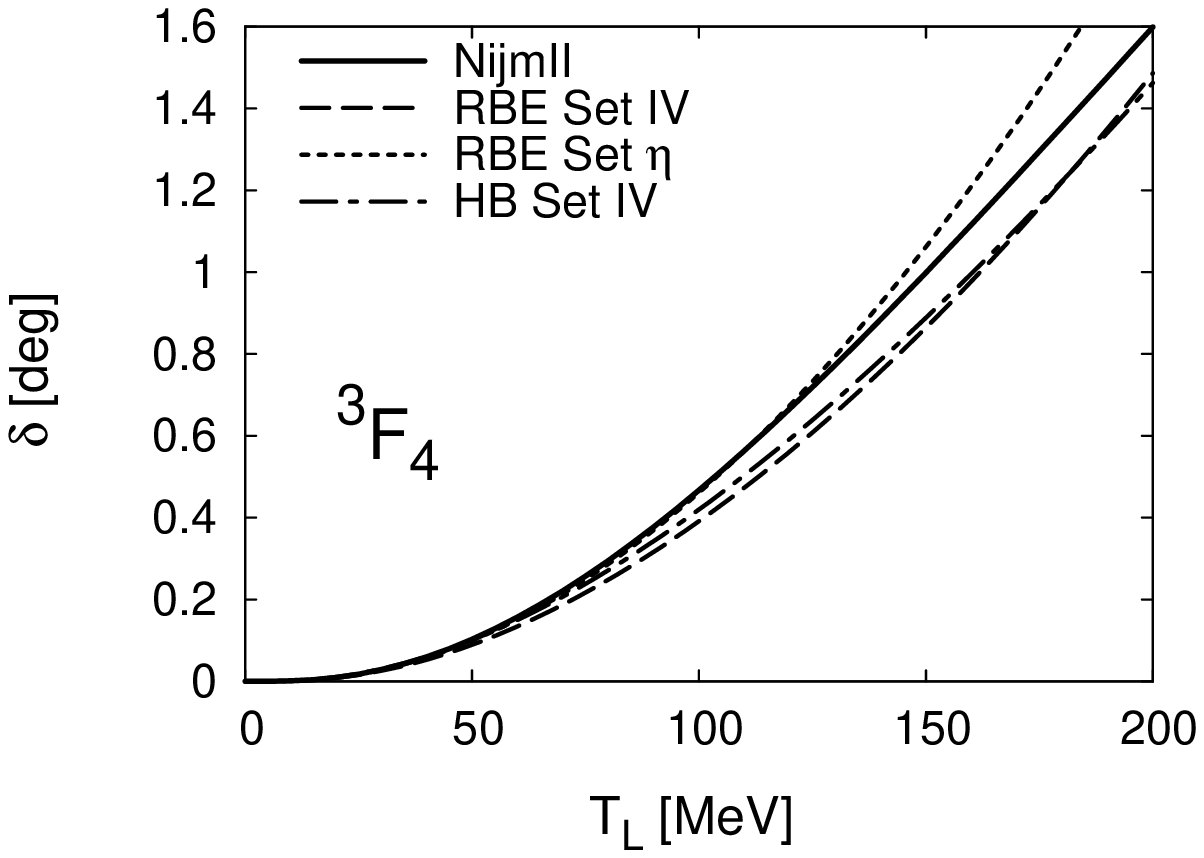, height=4cm, width=5cm}
\epsfig{figure=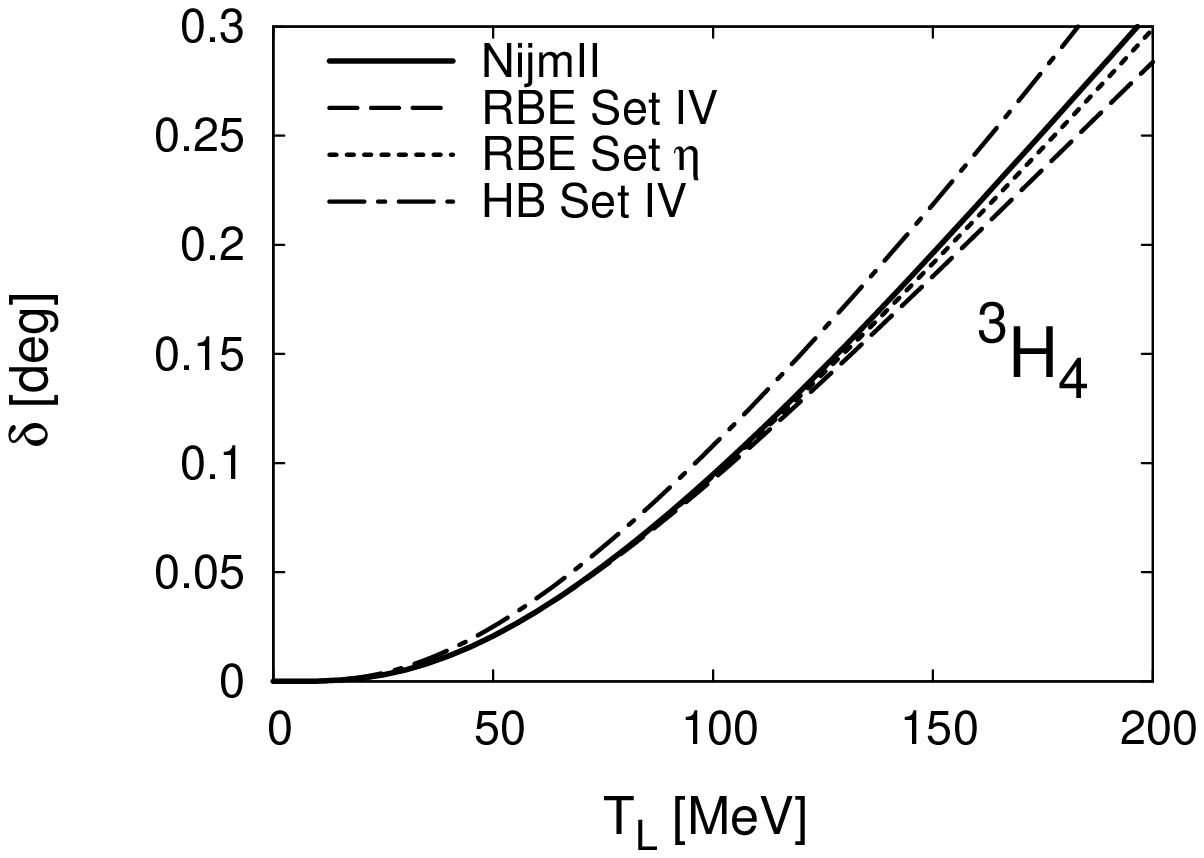, height=4cm, width=5cm}
\epsfig{figure=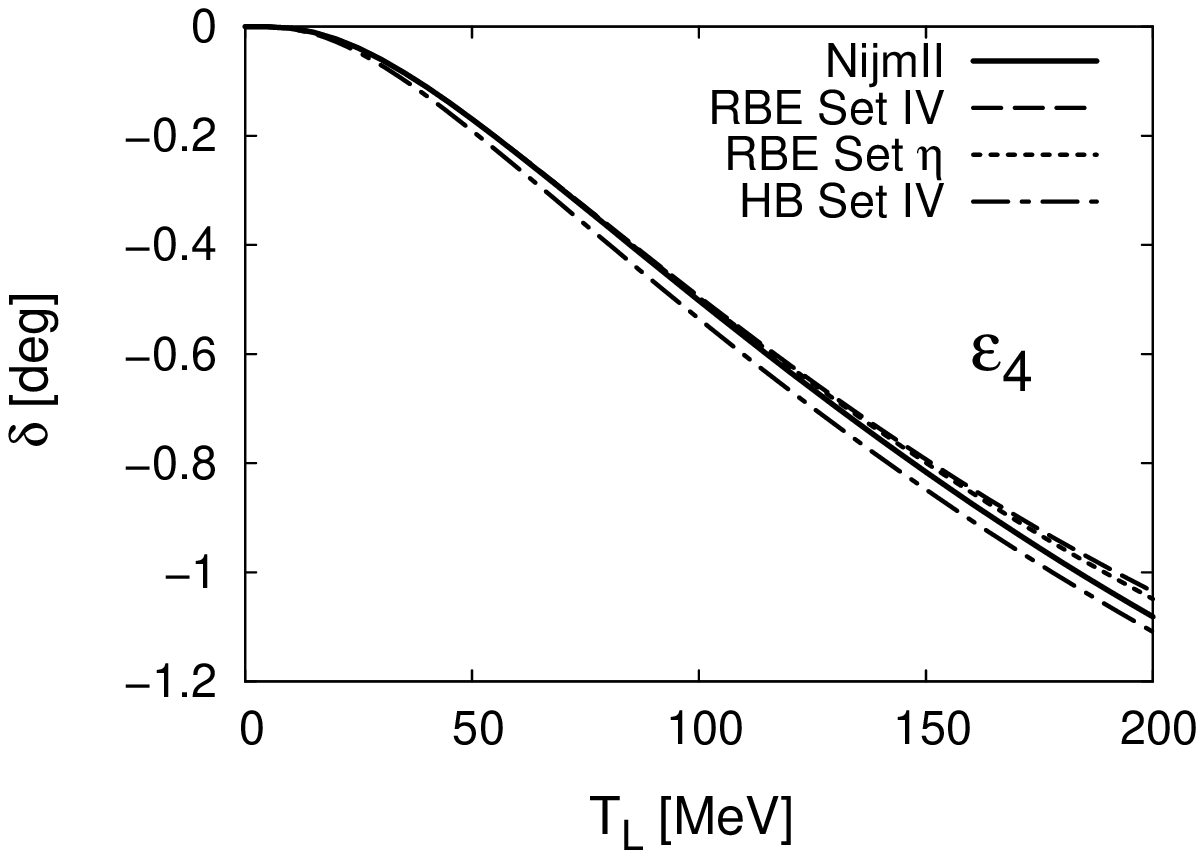, height=4cm, width=5cm} \\ 
\epsfig{figure=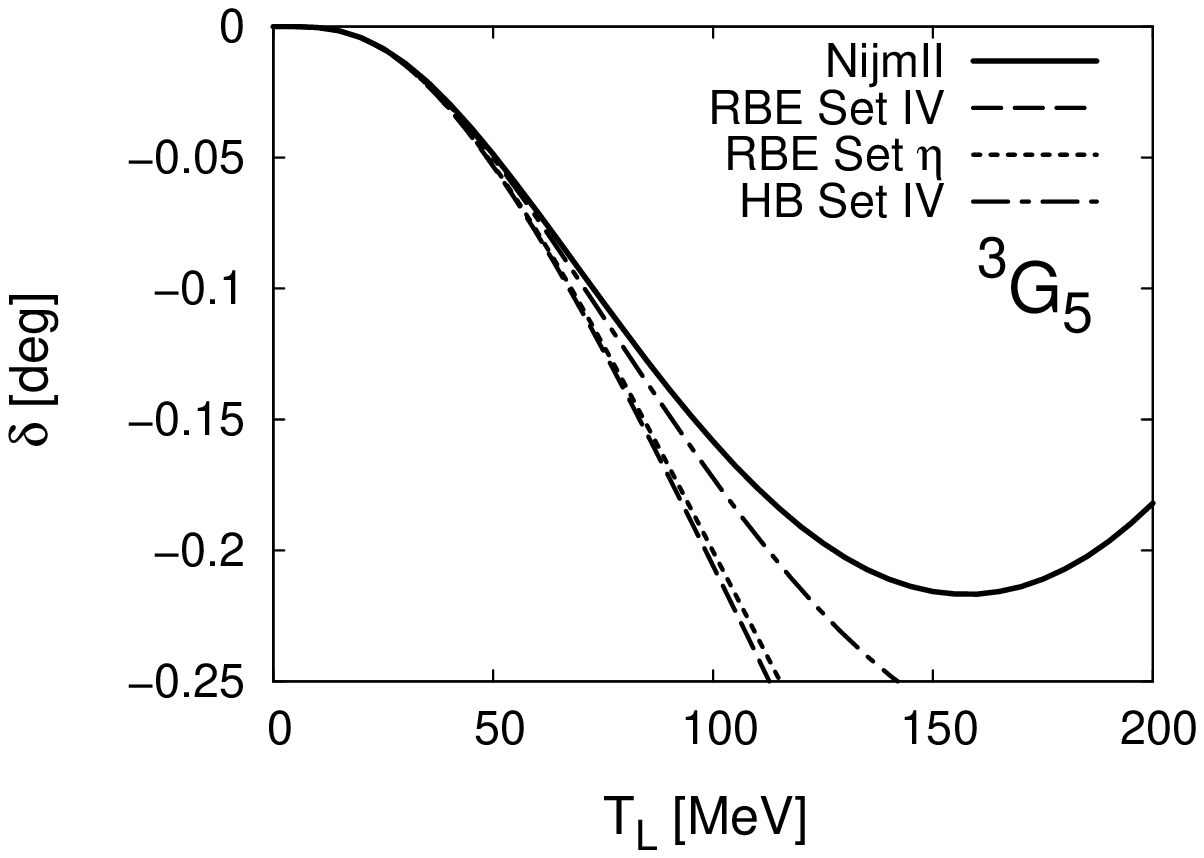, height=4cm, width=5cm}
\epsfig{figure=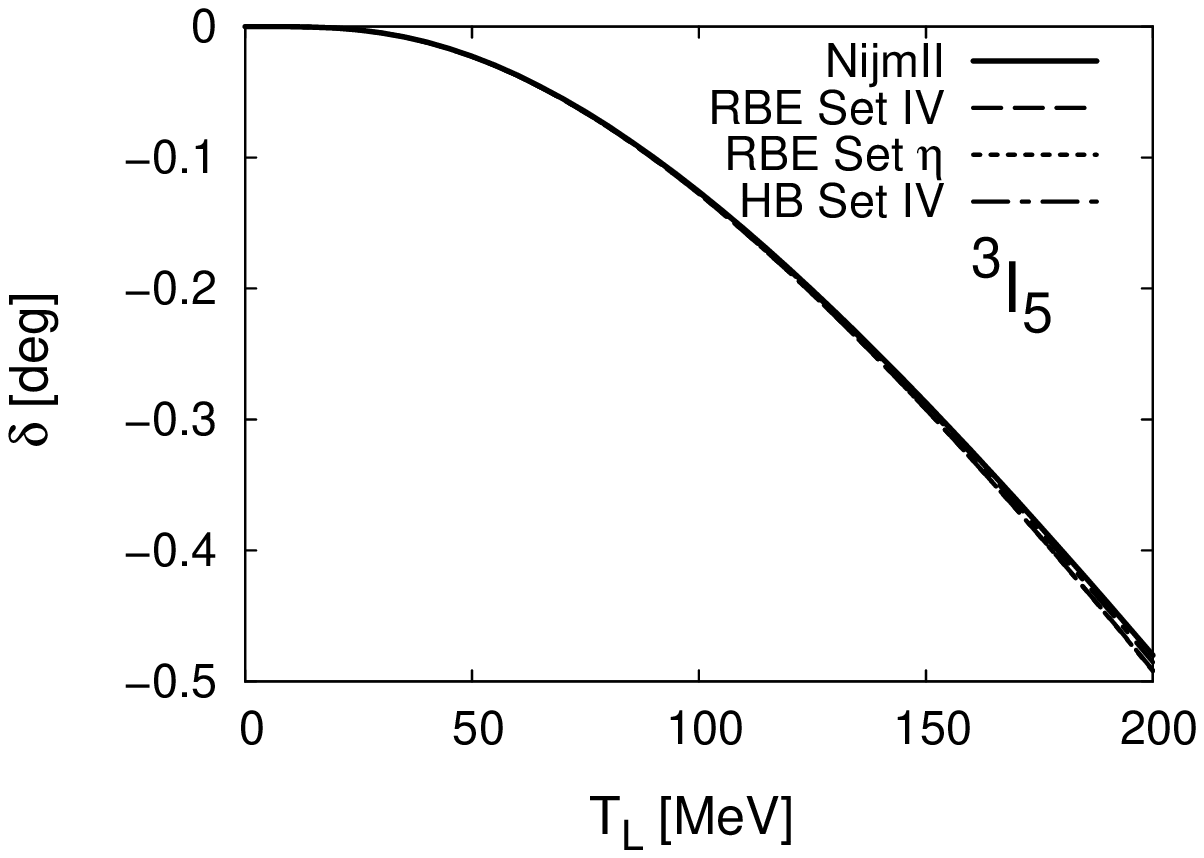, height=4cm, width=5cm}
\epsfig{figure=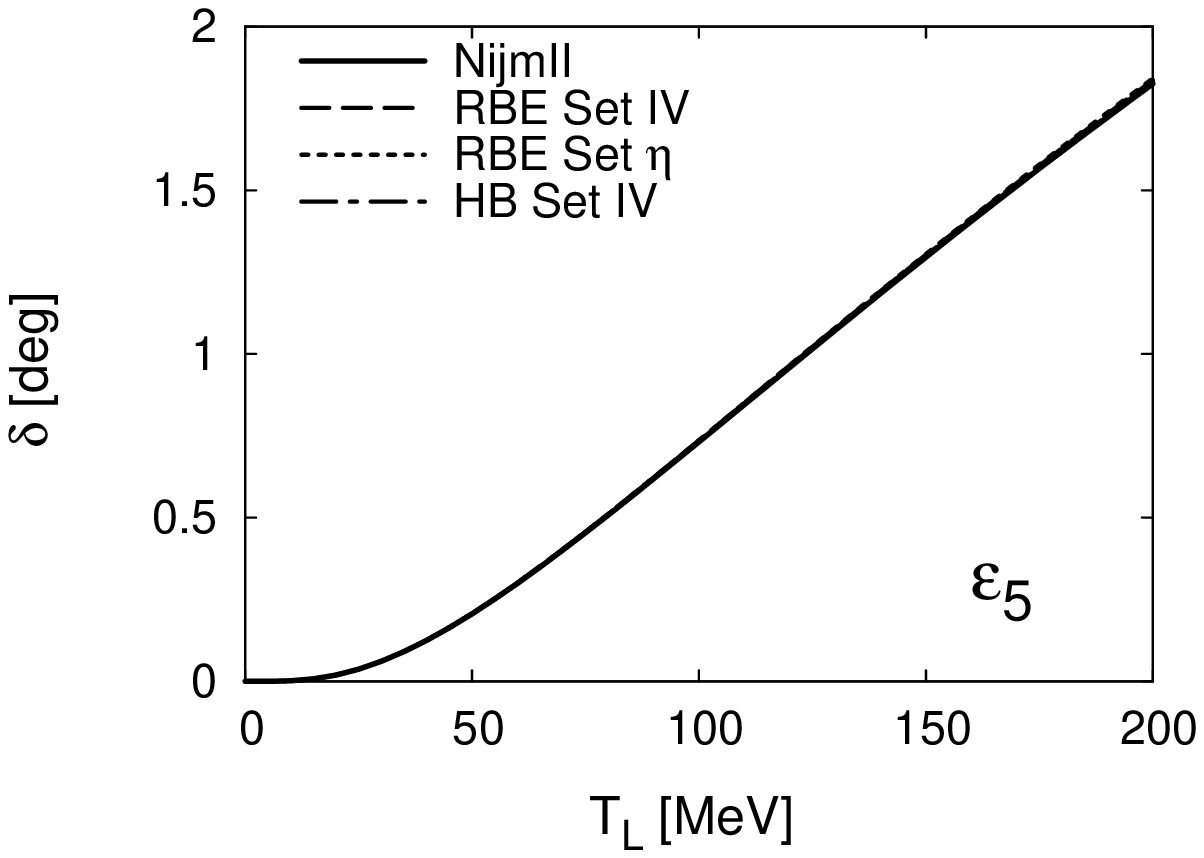, height=4cm, width=5cm}
\end{center}
\caption{np (SYM-nuclear bar) Coupled Spin Triplet Phase shifts for
the total angular momentum $j=0,1,2,3,4,5$ for relativistic baryon
expansion (RBE) and for the heavy baryon expansion (HBE) as a function
of as a function of the LAB energy compared to the Nijmegen partial
wave analysis~\cite{Stoks:1993tb,Stoks:1994wp}.}
\label{fig:cmp-coupled}
\end{figure*}

\section{Phase Shifts} 
\label{sec:phases}

We come to the calculation of the np phase shifts. In practice, this
requires a careful wave by wave study of the renormalized limit. As
can be seen in table~\ref{tab:table2}, all coupled triplet channels
have one attractive and one repulsive short distance $1/r^7$
eigenpotential. On the other hand, almost all singlet and uncoupled
triplet channels develop an attractive $1/r^7$ singularity at short
distances (sometimes depending on the parameter values). The only
exceptions we found are the $^1P_1$ and $^3P_0$ channels, the latter
depending on the precise values of the $c_{1,3,4}$ constants of the
chiral potential. This fact determines not only the number of
counterterms, but also the convergence pattern towards the
renormalized result. It reaches stability for cut-offs ranging in the
region $r_c=0.3-0.5 {\rm fm}$, depending on the particular partial
wave and also on the energy (see the discussion in the previous
section \ref{sec:deuteron} and Appendix~\ref{finite-cut-off}). Phase
shifts in coupled channels with one repulsive singular component have
been computed with either auxiliary boundary conditions $u_{0,j,l=j-1}
(r_c)=0$ and $u_{0,j,l=j-1}'(r_c)=0$ for zero energy states and
subsequent orthogonalization of the finite energy states by using a
complementary boundary condition as described in detail in
Ref.~\cite{PavonValderrama:2005gu} for the OPE case. It is important
to realize that even though renormalization requires in principle to
pursue the mathematical limit $r_c \to 0$, convergence is achieved in
practice by length scales which are not unrealistically small and in
fact are rather reasonable. This can be seen in
Fig.~\ref{fig:ps-running} where, for illustration purposes, some
selected low phases are depicted as a function of the short distance
cut-off for fixed laboratory energy values. As one generally expects
smaller values of $r_c$ are needed as the energy is increased. The
approach towards the renormalized value for any partial wave depends
on the attractive/repulsive character of the singular potential at
short distances. So, the $^1S_0$ and $^3P_0$ channels are purely
attractive and hence the finite cut-off corrections are ${\cal O} (
r_c^{\frac72+1}) $ up to oscillations~\cite{Valderrama:2007nu}. On the
other hand, the $^1P_1$ channel provides a repulsive case and hence
finite cut-off corrections are $ {\cal O} ( e^{-r_c^{-\frac52}}) $.
We remind that the smallest de Broglie wavelength
probed in NN interaction below pion production threshold is $\lambda
\sim 0.5 {\rm fm}$. Thus, the RB-potential and the present
renormalization construction also implement the desirable {\it a
priori} requirement that short distance details are indeed irrelevant
for the description of low energy properties.

In Figs.~\ref{fig:cmp-uncoupled} and \ref{fig:cmp-coupled} we present
the np (SYM-nuclear bar) renormalized phase shifts for the total
angular momentum $j=0,1,2,3,4,5$ for Spin Singlet and Uncoupled Spin
Triplet and Coupled Spin Triplet channels respectively. There we
compare the relativistic baryon expansion (RBE) and the heavy baryon
expansion (HBE) as a function of the LAB energy compared to the
Nijmegen partial wave analysis~\cite{Stoks:1993tb,Stoks:1994wp}. For
definiteness we use the chiral constants $c_1$, $c_3$ and $c_4$ of
Ref.~\cite{Entem:2003ft} (Set IV), which already provided a good
description of deuteron properties after
renormalization~\cite{Valderrama:2005wv} at NNLO. This choice allows a
more straightforward comparison to the N$^3$LO calculation of
Ref.~\cite{Entem:2003ft} with finite cut-offs. We also compare with
the Set $\eta$ which takes to the same value of $c_1$ and the
readjusted values $c_3=-3.8 {\rm GeV}^{-1}$ and $c_4 = 4.5 {\rm
GeV}^{-1}$ based on our improved description of the deuteron in
Sec.~\ref{sec:deuteron}.  Unless otherwise stated, the needed low
energy parameters for these figures are {\it always} taken to be those
of Ref.~\cite{Valderrama:2005ku} for the NijmII potential (see
Table~\ref{tab:table2}). As can be clearly seen, the RB-TPE with this
Set-$\eta$ not only improves deuteron properties but also the phase
shifts all over. 
Again, one should not overemphasize this agreement, but it is
rewarding to see that there is a general trend to stability and
improvement in some channels (such as $^1D_2$,$^3P_1$, $^3P_0$) when
the RB-TPE potential is considered, while the quality of description
is not worsened in other channels. At the same time one should stress,
however, that generally speaking this potential needs much less
counter-terms as the corresponding HB counterpart (about a
half). Actually, in the Lorentz-invariant potential case one has at
most one free parameter for channel as compared to the three
parameters for channels in the coupled triplets. This is due to the
attractive-repulsive short distance character of the coupled channel
RB-TPE as compared to the attractive-attractive HB-TPE potential in
these coupled channels. Again, we remind that the RB-TPE provides the
correct analytic behavior of the exchange of two pions at large
distances when $\Delta$ and other excitations are not explicitly
considered.

\section{Conclusions} 
\label{sec:concl}

In the present paper we have analyzed the renormalization of all
partial waves for NN scattering and the bound deuteron state for the
Chiral Two Pion Exchange Potential computed in a relativistic baryon
Expansion. Our main motivation has been to consider a potential where
the asymptotic TPE effects are consistently taken care of.  This gives
us some confidence that long distance physics is faithfully
represented by a common exponential fall-off factor $e^{- 2 m_\pi
r}$. At the level of calculation considered here $\Delta$ and other
excitations are not explicitly taken into account; within a
relativistic baryon expansion the contribution of this degree of
freedom to the NN potential has never been computed.  In addition, in
the relativistic baryon expansion there appear non-local terms in the
potential which involve the sum of initial and final momentum space
operators which fortunately turn out to be rather small all over the
range as compared to the local contributions to the potential. This
simplifies the analysis tremendously since standard coordinate space
methods can be used to solve the non-perturbative scattering problem
and deuteron bound state and their subsequent and necessary
renormalization. As we have repeatedly stressed along the paper this
Lorentz-invariant potential presents a $1/r^7$ singularity at the
origin which demands renormalization in order to get a finite and
unique result when the TPE potential is assumed to be valid all over
the range from the origin to infinity. This can be done by introducing
a number of (potential independent) counter-terms and consequently
physical renormalization conditions must be specified. In practice
they are fixed to the values of threshold parameters, mainly
scattering lengths at zero energy. Actually, we have noted that the
number of necessary counter-terms is drastically reduced when the
Lorentz-invariant baryon potential is compared to the heavy baryon
expanded TPE potential, while both potentials are specified by the
same parameters. Thus, less input is needed to predict the NN
phase-shifts. Although the precise number of counter-terms depends on
the parameters of the potential we find that, typically, for the
channels with total angular momentum $j=0$ to $5$ we need 13 in the
Lorentz-invariant case as compared to about 27 in the HB
potential. Actually, it is noteworthy that with about a half of the
counter-terms the overall agreement is improved. This is particularly
striking in the $^3P_0$ and the $^3S_1-^3D_1$ (deuteron) channels. In
other channels the improvement is moderate, indicating missing shorter
range contributions to Eq.~(\ref{eq:full_pot}). Although a deeper
understanding on why this dramatic reduction of the number of
counter-terms happens would very helpful, and we have not attempted a
large scale fit, it is very rewarding that the implementation of the
correct and fairly complete long range physics deduced from One and
Two Pion Exchange in conjunction with the requirement of
renormalizability provides a rather reasonable description of the NN
scattering data below pion production threshold.

\begin{acknowledgments}

We would like to thank the hospitality of the European Center for
Theoretical Studies on Nuclear Physics and Related Areas (ECT*) in
Trento, where this work was initiated.  M.P.V. also thanks
W. Broniowski and P. Bo{\.z}ek for their kind hospitality in Krakow
where part of this work has been carried out.  

The work of M.P.V. and E.R.A.
is supported in part by funds provided by the Spanish DGI and FEDER
funds with grant no. FIS2005-00810, Junta de Andaluc{\'\i}a grants no.
FQM225-05, EU Integrated Infrastructure Initiative Hadron Physics
Project contract no. RII3-CT-2004-506078.  The work of R.H. was
supported by DOE Contract
No.DE-AC05-06OR23177, 
under which SURA operates the Thomas Jefferson National Accelerator 
Facility, and by the BMBF under contract number 06BN411. 

\end{acknowledgments}

\appendix 

\section{Analytical determination of finite cut-off corrections}
\label{finite-cut-off} 

In this appendix we determine the finite cut-off corrections to
renormalized deuteron properties when the the two coupled channel
potential is so that there is one attractive and one repulsive short
distance eigenpotential. This is the case of OPE and RB-TPE potentials
discussed in this paper. In this case it is simplest to discuss the
auxiliary boundary condition
\begin{eqnarray} 
u(r_c)=0
\label{bc_short} 
\end{eqnarray}  
The analysis of other auxiliary boundary  conditions such as $u'(r_c)=0$ which
has actually been used in the numerical calculations is a bit messier
but the final conclusion is essentially the same as with
Eq.~(\ref{bc_short}).

From the superposition principle of boundary conditions we may write 
\begin{eqnarray}
u(r) &=& u_S (r) + \eta \, u_D (r) \, , \nonumber \\  
w(r) &=& w_S (r) + \eta \, w_D (r) \, ,
\end{eqnarray} 
where $(u_S(r),w_S(r))$ and $(u_D(r),w_D(r))$ solve the deuteron
problem, Eq.~(\ref{eq:sch_coupled}), with the long distance boundary
conditions Eq.~(\ref{eq:bcinfty_coupled}) when taking $(A_S,A_D)=(1,0)$
and $(A_S,A_D)=(0,1)$ respectively. On the other hand, at short
distances the coupled channel potential is diagonalized by an
orthogonal transformation 
\begin{eqnarray}
{\bf G} = \left( \begin{matrix} \cos \theta & \sin \theta \\ -\sin \theta & 
\cos \theta \end{matrix} \right) 
\end{eqnarray} 
where the mixing angle $\theta$ depends on the parameters of the
potential only. For instance, for OPE one has $\cos \theta =-1/
\sqrt{3} $, see Ref.~\cite{PavonValderrama:2005gu}. The reduced
potential behaves as
\begin{eqnarray}
{\bf U} (r) \to \frac1{r^n} \, \, {\bf G} \left( \begin{matrix}
+R_+^{n-2} & 0 \\ 0 & -R_-^{n-2} \end{matrix} \right) {\bf G}^{-1}
\end{eqnarray} 
where $R_+$ and $R_-$ are the Van der Waals length scales associated
to the repulsive and attractive channels respectively. Defining the
general short distance solutions of the decoupled problems
\begin{widetext} 
\begin{eqnarray}
v_{i,+} (r) & = & \left(\frac{r}{R_+}\right)^{n/4}  \left\{ a_{i,+} 
\exp \left[- \frac{2}{n-2}
\left(\frac{R_+}{r}\right)^{\frac{n-2}2} \right]   +  b_{i,+} 
\exp \left[+ \frac{2}{n-2}
\left(\frac{R_+}{r}\right)^{\frac{n-2}2} \right] \right\} \nonumber \\ 
v_{i,-} (r) & = & \left(\frac{r}{R_-}\right)^{n/4}  c_{i,-} 
\sin \left[ \frac{2}{n-2}
\left(\frac{R_-}{r}\right)^{\frac{n-2}2} + \varphi_i \right]  
\label{eq:+-short}
\end{eqnarray} 
\end{widetext} 
where $i=S,D$ and $a_{i,+}$, $b_{i,+}$, $c_{i,-}$ , $\varphi_{i}$
are fixed constants which depend on the potential only and may be
determined by integrating in from long distances the asymptotic wave
functions $(u_S(r),w_S(r))$ and $(u_D(r),w_D(r))$.  We remind that
$n=7$ for the RB-TPE potential while $n=3$ for the OPE potential.
Note that we must include here also the diverging exponential at the
origin for the repulsive eigenpotential. The
solutions at short distances behave as
\begin{eqnarray}
\left( \begin{matrix} u_i(r) \\ w_i (r) \end{matrix} \right)  \to 
\left( \begin{matrix} \cos \theta & \sin \theta \\ -\sin \theta & 
\cos \theta \end{matrix} \right) \left( \begin{matrix} v_{i,+}(r) \\ v_{i,-} (r) \end{matrix} \right)   
\end{eqnarray} 
where $i=S,D$. Thus, using the short distance boundary condition,
Eq.~(\ref{bc_short}) which selects the regular solution at the origin
and eventually kills the diverging exponentials when $r_c \to 0$, and
gathering all subsequent equations we get
\begin{eqnarray}
\eta (r_c) &=& -  \frac{u_S (r_c)}{u_D(r_c)}  \nonumber \\ 
& \to & \frac
{\cos \theta \, v_{S,+} (r_c) + \sin \theta \, v_{S,-} (r_c) }
{\cos \theta \, v_{D,+} (r_c) + \sin \theta \, v_{D,-} (r_c) }
\end{eqnarray} 
The limiting value is controlled by the short distance diverging
exponentials in Eq.~(\ref{eq:+-short}) and is given by
\begin{eqnarray}
\eta (0) = -\frac {b_{S,+}} {b_{D,+}}
\end{eqnarray} 
Deviations from this value for small cut-offs $r_c$ can be directly
determined from Eq.~(\ref{eq:+-short}) yielding
\begin{widetext} 
\begin{eqnarray}
\frac{\eta (r_c)}{\eta(0)} =  1 + \tan \theta \left(\frac{R_+}{R_-} \right)^\frac{n}4  \left[ \frac{c_{S,-}}{b_{S,+}}-\frac{c_{D,-}}{b_{D,+}} \right] 
\exp \left[- \frac{2}{n-2}
\left(\frac{R_+}{r_c}\right)^{\frac{n-2}2} \right] 
\sin \left[ \frac{2}{n-2}
\left(\frac{R_-}{r_c}\right)^{\frac{n-2}2} + \varphi_i \right] + \dots 
\end{eqnarray} 
\end{widetext} 
showing that, up to oscillations, finite cut-off corrections in the
deuteron are $ {\cal O} ( e^{-r_c^{-\frac12}}) $ for OPE and $ {\cal
O} ( e^{-r_c^{-\frac52}}) $ for RB-TPE. The generalization to other
auxiliary boundary conditions and other deuteron properties is
straightforward with identical result in the order of finite cut-off
effects. 

The case of scattering states is more tedious and will not be
discussed in detail here but can also be analyzed with a combination
of the coupled channel formulas of Ref.~\cite{Valderrama:2007nu} (see
Sect. V of that paper) and the bound state results of the present
appendix. Finite cut-off effects for the S-matrix scale similarly as
in the bound state case, i.e.  up to oscillations they are $ {\cal O}
( e^{-r_c^{-\frac12}}) $ for OPE and $ {\cal O} ( e^{-r_c^{-\frac52}})
$ for RB-TPE.


\end{document}